\pdfoutput=1

\documentclass[11pt]{article}
\usepackage[preprint]{acl}

\usepackage{lineno}
\usepackage{times}
\usepackage{latexsym}
\usepackage{longtable}

\usepackage[T1]{fontenc}

\usepackage[utf8]{inputenc}

\usepackage{microtype}

\usepackage{inconsolata}

\usepackage[noend]{algorithmic}
\usepackage{amsfonts}       
\usepackage{amsmath,amssymb}        
\usepackage{arydshln}

\usepackage{hyperref}       
\usepackage{url}            
\usepackage{booktabs}       
\usepackage{listings}
\usepackage{nicefrac}       
\usepackage{makecell}

\usepackage{enumitem}
\setlist[itemize]{align=parleft,left=0.5em..1.em}

\usepackage{multicol}
\usepackage{multirow}
\usepackage{tabularx}
\usepackage{float}
\usepackage{cleveref}
\usepackage{titletoc}
\usepackage{graphicx}
\usepackage{color, colortbl}
\usepackage{wrapfig}
\usepackage{dblfloatfix}
\usepackage{framed}
\usepackage{xcolor}         
\definecolor{darkblue}{HTML}{000099} 
\definecolor{red-color}{RGB}{238, 117, 120}
\definecolor{green-color}{RGB}{112, 173, 71}
\definecolor{blue-color}{RGB}{91, 155, 213}
\definecolor{orange-color}{RGB}{237, 125, 49}
\definecolor{dkgreen}{rgb}{0,0.6,0}
\definecolor{LightGrey}{HTML}{d9d9d9}
\definecolor{mauve}{rgb}{0.58,0,0.82}

\hypersetup{colorlinks=true, citecolor=darkblue, linkcolor=darkblue, urlcolor=darkblue}

\title{TreeHop: Efficient Embedding-Level Query Rewriter}

\author{Zhonghao Li\textsuperscript{\rm 1}, Kunpeng Zhang\textsuperscript{\rm 1}, Jinghuai Ou\textsuperscript{\rm 3}, Shuliang Liu\textsuperscript{\rm 2,3}, Xuming Hu\textsuperscript{\rm 2,3} \\
\\
\textsuperscript{\rm 1}University of Maryland \\
\textsuperscript{\rm 2}The Hong Kong University of Science and Technology (Guangzhou) \\
\textsuperscript{\rm 3}The Hong Kong University of Science and Technology \\
\texttt{al1231@terpmail.umd.edu}, \texttt{kpzhang@umd.edu} \\
}

\begin{document}

\maketitle

\begin{abstract}
Retrieval-augmented generation (RAG) systems face significant challenges in multi-hop question answering (MHQA), where complex queries require synthesizing information across multiple document chunks. Existing approaches typically rely on iterative LLM-based query rewriting and routing, resulting in high computational costs due to repeated LLM invocations and multi-stage processes. To address these limitations, we propose TreeHop, an embedding-level framework without the need for LLMs in query refinement. TreeHop dynamically updates query embeddings by fusing semantic information from prior queries and retrieved documents, enabling iterative retrieval through embedding-space operations alone. This method replaces the traditional "Retrieve-Rewrite-Vectorize-Retrieve" cycle with a streamlined "Retrieve-Embed-Retrieve" loop, significantly reducing computational overhead. Moreover, a rule-based stopping criterion is introduced to further prune redundant retrievals, balancing efficiency and recall rate. Experimental results show that TreeHop rivals advanced RAG methods across four open-domain MHQA datasets, achieving comparable performance with only 2.2\%-29.4\% of the parameter size of concurrent solutions and reducing the query latency by 92.8\%-97.8\%. This makes TreeHop a faster and more cost-effective solution for low-resource or latency-sensitive deployment. For reproducibility purposes, codes and data are available here\footnote{\url{https://github.com/allen-li1231/TreeHop-RAG}}.

\end{abstract}

\section{Introduction}

Large Language Models (LLMs)~\citep{deepseekai2025deepseekr1incentivizingreasoningcapability,bai2025qwen3vltechnicalreport} understand queries~\citep{ouyang2022training,brown2020language} and generate fluent text, yet on domain-specific~\citep{li2024refinerrestructureretrievalcontent,zhang2024raft} or knowledge-intensive~\citep{kandpal2023large} tasks they hallucinate~\citep{zhang2023sirens} once a query exceeds their parametric knowledge~\citep{muhlgay2024generating}. Retrieval-Augmented Generation (RAG)~\citep{lewis2021retrievalaugmented, gao2024retrievalaugmented} addresses this by retrieving relevant chunks from external knowledge bases, producing more faithful~\citep{khandelwal2020generalization} and generalizable~\citep{kamalloo2023evaluating} answers.

However, the conventional single-retrieval paradigm of RAG falters in multi-hop question answering (MHQA) scenarios~\citep{yang2018hotpotqa,xanh2020_2wikimultihop,tang2024multihopragbenchmarkingretrievalaugmentedgeneration,trivedi-etal-2022-musique}, where answers require synthesizing information from multiple document chunks. For instance, consider the query "\textit{Who is the grandfather of Donald Trump?}" A single retrieval might return a chunk stating "\textit{Donald John Trump was born on June 14, 1946..., the fourth child of Fred Trump and Mary Ann Macleod Trump.}", but resolving the grandfather requires a follow-up query like "\textit{Who is the father of Fred Trump?}". This typical multi-hop scenario reveals the need to dynamically compose new query based on information in relevant document chunk\footnote{For more discussion about the need of query rewriters for MHQA task, please refer to \autoref{why_need_query_rewriter_for_MHQA}.}. Query-rewriters~\citep{ma2023queryrewritingretrievalaugmentedlarge}, routers~\citep{zhuang2024efficientragefficientretrievermultihop} and iterative loops~\citep{shao2023enhancingretrievalaugmentedlargelanguage} address this by refining queries with retrieved information, but the repeated LLM invocations impose substantial latency.

\begin{figure*}[t]
    \vspace*{-10mm}
    \centering
    \includegraphics[width=\linewidth]{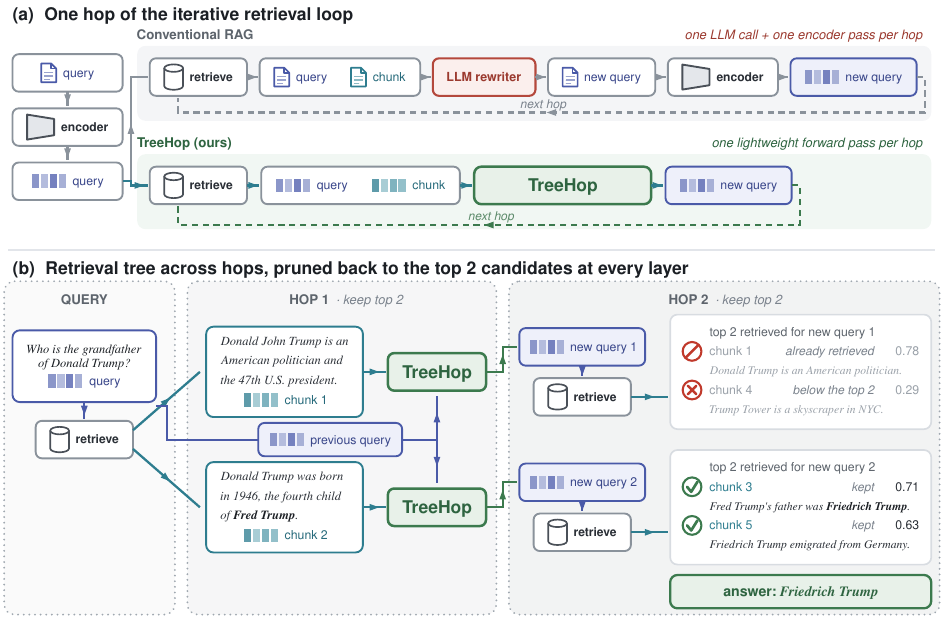}
    \caption{TreeHop replaces LLM-based query rewriting with an update carried out directly on embeddings. \textbf{(a)} Conventional iterative RAG rewrites the query with an LLM and re-embeds it at every hop; TreeHop instead fuses the previous query embedding with the retrieved chunk embedding, collapsing ``Retrieve-Rewrite-Vectorize-Retrieve'' into ``Retrieve-Embed-Retrieve''.
    \textbf{(b)} The question is encoded once by embedding model. 
    Every retrieved chunk embedding spawns its own next-hop query via TreeHop model; the candidates that all branches retrieve at a layer are ranked together and cut back to $K$ (2 shown in the figure) by \emph{redundancy pruning} and \emph{layer-wise top-$K$ pruning}.}
    \label{fig:treehop_iteration}
    \vspace*{-4mm}
\end{figure*}

To address these limitations, we propose TreeHop, a framework enabling embedding-level query updates without requiring LLM rewrites.
Inspired by the semantic and structural properties of sentence embeddings~\citep{zhu-etal-2018-exploring}, TreeHop dynamically predicts next-step query embeddings by fusing prior queries and retrieved content embeddings (\autoref{fig:treehop_iteration}b).
This approach collapses the traditional "Retrieve-Rewrite-Vectorize-Retrieve" cycle (\autoref{fig:treehop_iteration}a) into a streamlined "Retrieve-Embed-Retrieve" loop, significantly reducing computational costs.
TreeHop further introduces two pruning strategies to ensure computational efficiency: \textit{redundancy pruning} terminates paths where the retrieved chunks have been seen in previous iterations, while \textit{layer-wise top-K pruning} retains only the top-ranked retrieval candidates at each step, curbing exponential branch growth.

TreeHop employs a gated cross-attention mechanism~\citep{vaswani2023attentionneed} to effectively focus on extracting salient information from retrieved chunks, making the model effective while parameterizing with only 0.18 billion parameters. Trained with contrastive learning~\citep{chen2020simpleframeworkcontrastivelearning, wu-etal-2022-infocse}, TreeHop is capable of achieving a performance comparable to computationally intensive multi-hop methods across four benchmarks while maintaining significantly lower latency.
Remarkably, TreeHop reduces retrieval latency by 92.8\%-97.8\% compared to LLM-based methods while sacrificing only 0.9\% of the recall rate at maximum, even outperforming some advanced systems by 9.1\% in deeper retrieval iterations.

In summary, our work makes three key contributions to the class of efficient, iterative retrieval frameworks:
\begin{itemize}
\item A novel embedding-updating mechanism that replaces LLM-driven iterative query rewrites with lightweight neural operations, enabling linear computational complexity for MHQA tasks.
\item An efficient rule-based stopping criterion that controls branching factor growth while maintaining performance in retrieval iteration.
\item Empirical validation demonstrating superior efficiency-accuracy trade-offs in four MHQA tasks. Our approach bridges the gap between computational efficiency and retrieval effectiveness, offering a scalable solution for diverse knowledge-intensive applications.
\end{itemize}

\section{Preliminaries}

\subsection{Multi-hop Retrieval-Augmented Generation}
Multi-hop RAG iterates retrieval. Iter-RetGen~\citep{shao2023enhancingretrievalaugmentedlargelanguage} feeds the LLM's partial answer back as context for the next round, and query rewriting reformulates the query from retrieved context~\citep{ma2023query, vakulenko2020question}; both re-generate and re-encode text at every hop. EfficientRAG~\citep{zhuang2024efficientragefficientretrievermultihop} trims this with a trained filter and rewriter. Two recent MHQA systems pursue our objective from the opposite cost regime: Q-DREAM~\citep{ye-etal-2025-optimizing} also optimises the question semantic space, but through a decomposition model, a dependency optimiser and cluster-wise LoRA tuning over 7B backbones (roughly 4s per query); ChainRAG~\citep{zhu-etal-2025-mitigating} identifies ``lost-in-retrieval'', where key entities dropped during sub-question decomposition break the reasoning chain, and repairs it by prompting an LLM against a sentence graph. Each corroborates that recovering the bridge entity is the crux of multi-hop retrieval, the premise TreeHop builds on, yet each spends an LLM generation per hop to act on it. RAPTOR~\citep{sarthi2024raptor} moves that cost off the query path, clustering and summarising passages into a tree at index time so one retrieval can reach evidence that would otherwise take several hops.
To our knowledge, EfficientRAG is the only efficient iterative retrieval model with open-source parameters; since our goal is efficiency, we treat accuracy-focused methods—such as the model-agnostic Iter-RetGen and the tree-structured RAPTOR—as complementary rather than as baselines.

\subsection{Sentence Representation Learning and Contrastive Learning}
Sentence representation learning encodes a sentence into a single fixed-dimensional embedding, and contrastive objectives are the standard way to train it. SimCSE~\citep{gao-etal-2021-simcse} employs InfoNCE~\citep{oord2019representationlearningcontrastivepredictive} to pull two encodings of the same sentence together while pushing apart encodings of different sentences:
\begin{equation}
    \mathcal{L}_{\rm{infoNCE}} = - \sum_{i=1}^N\log \frac{e^{sim(f(x_i),f(x_i)')/\tau}}{\sum_{j=1}^Ne^{sim(f(x_i),f(x_j)')/\tau}},
\label{eq:contrastive}
\end{equation}
where $N$ is the batch size, $\tau$ a temperature hyperparameter, $sim(f(x_i),f(x_j)')=\frac{f(x_i)\top f(x_j)'}{\Vert f(x_i) \Vert \cdot \Vert f(x_j)' \Vert}$ the cosine similarity, $f(\cdot)$ the encoder, and $(x_i, x'_i)$ a semantically related pair from the positive set $\mathcal{D} = \{(x_i, x'_i) \}_{i=1}^m$.



\begin{figure*}
    \vspace*{-10mm}
    \centering
    \includegraphics[width=\linewidth]{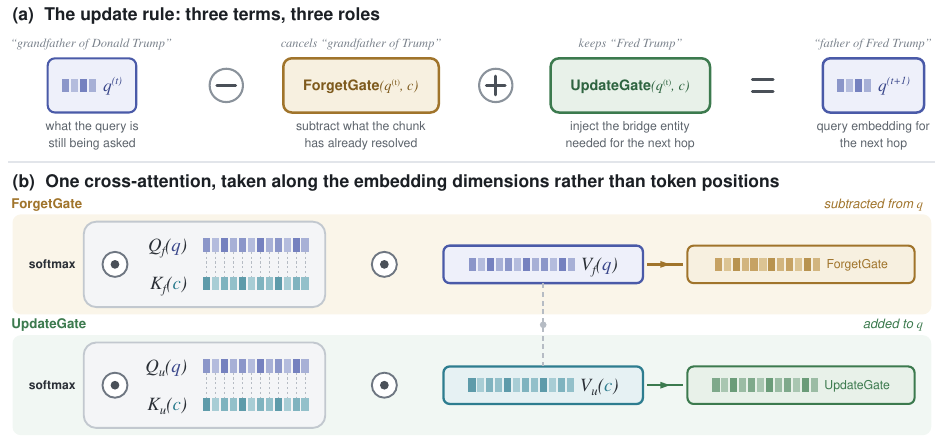}
    \caption{The TreeHop update. \textbf{(a)} The next-hop query embedding is assembled from the three terms of \autoref{eq:treehop_architecture1}, glossed on the running example of \autoref{fig:treehop_iteration}. \textbf{(b)} The two gates are the same cross-attention and differ only in the source of their values; \autoref{sec:model_architecture} explains why the attention runs along the embedding dimensions.}
    \label{fig:treehop_architecture}
    \vspace*{-4mm}
\end{figure*}

\section{The Proposed Method: TreeHop}
TreeHop generates the next query embedding by integrating the previous query embedding with retrieved content embeddings, in an architecture optimised for retrieval performance at a compact parameter size.

\subsection{Problem Formulation}
At retrieval iteration step $r$, given query embedding $q$, a set of the top $K$ document chunk embeddings, $\mathcal{T} = \{c^i \}_{i=1}^K$, is retrieved using the retriever $g(q_r, K)$. The TreeHop model then produces the corresponding query embedding set $\mathcal{Q}_1 = \{q_{r+1}^i\}_{i=1}^K$ for the subsequent $r+1$ step.


\begin{equation}
    q_{r+1}^i = \text{TreeHop}(q_r, c_r^i), c_r^i \in \mathcal{T}_r
\end{equation}

Note, that the TreeHop framework defaults to the base retriever under the single-hop retrieval scenario, as no iterative query rewrite is needed.
Please refer to \autoref{fig:treehop_iteration}b for an illustration of TreeHop inference with the stopping criterion described in \autoref{sec:stop_criterion}.

\subsection{Model Architecture}
\label{sec:model_architecture}
The core of TreeHop’s query update is a pair of cross-attention gates, the $ForgetGate$ and the $UpdateGate$ (see \autoref{fig:treehop_architecture}b), which together modulate how information from prior queries $q$ and retrieved chunks $c$ is discarded or retained. The intuition behind is that we only need to remove information present in both embeddings, and update information yet to be further retrieved from the retrieved chunks to form a new query embedding.
\begin{equation} \label{eq:treehop_architecture1}
    \small
    \begin{aligned}
    \text{TreeHop}(q_r, c_r^i) = {} & q_r - ForgetGate(q_r, c_r^i) \\
    & + UpdateGate(q_r, c_r^i)
    \end{aligned}
\end{equation}

The $ForgetGate(q_r, c_r^i)$ term suppresses the semantics that the retrieved chunk has already resolved (\autoref{fig:treehop_architecture}a). Rather than subtracting the chunk embedding directly, it is designed to attend over the query-chunk pair and remove only those query dimensions the chunk accounts for, which prevents redundant retrieval of information already captured (e.g., if the chunk confirms "Fred Trump is Donald’s father," the model avoids re-searching for Fred Trump in subsequent hops).
The $UpdateGate(q_r, c_r^i)$ selectively incorporates new information from $c_r^i$ to form the next query. Both gates revise the cross-attention mechanism~\citep{vaswani2023attentionneed} to work on embedding level, and are identical in structure except for the source of their values: $ForgetGate$ draws values from the query $q$, whereas $UpdateGate$ draws them from the chunk $c$.

\begin{equation} \label{eq:treehop_architecture2}
    \small
    \begin{aligned}
    ForgetGate(q, c) &= CrossAttn_f(q, c) \\
     &= \text{softmax}(Q_f(q) \odot K_f(c)) \odot V_f(q) \\
    UpdateGate(q, c) &= CrossAttn_u(q, c) \\
     &= \text{softmax}(Q_u(q) \odot K_u(c)) \odot V_u(c)
    \end{aligned}
\end{equation}

Where $Q_f, K_f, V_f$ and $Q_u, K_u, V_u$ are weight and bias matrices learned separately for the two gates. Conventional cross-attention scores one token against another and normalizes the resulting map over sequence positions. That axis is unavailable to us: the frozen encoder emits a single vector per query and per chunk, so no token positions survive for the gates to attend over. We therefore replace the dot products with element-wise products and take the softmax along the $d$ embedding dimensions instead, so that each gate produces one weight per dimension which scales its value vector independently (\autoref{fig:treehop_architecture}b). Attention thus selects \emph{which coordinates of the embedding} carry the semantics to discard or to inject, which is precisely the granularity a query rewrite needs once the text has been pooled away into a single vector. See \autoref{subsec:ablation_architecture} for ablations and \autoref{intuition_behind_treehop_architecture} for the intuition behind the design.

\subsection{Stopping Criterion} \label{sec:stop_criterion}
TreeHop produces a query embedding for every retrieved chunk, so unchecked iteration grows the retrieved set exponentially ($O(n^r)$) without proportional gains in accuracy. We therefore introduce rule-based stopping criteria that prune redundant branches so that only promising paths advance.



\paragraph{Redundancy Pruning}
We terminate branches where the document chunk has been retrieved in the previous iterations, as depicted in line \autoref{eq:redundancy_pruning}, \autoref{fig:treehop_inference_step}.

\paragraph{Layer-wise Top-$K$ Pruning}
At each retrieval layer, we retain only the top-$K$ chunks with highest similarity scores across all generated query embeddings. This reduces the branching factor from $O(n^r)$ to $O(nr)$ by focusing computation on the most promising paths, as shown in line \autoref{eq:layerwise_top_pruning}, \autoref{fig:treehop_inference_step}. Please refer to \autoref{more_explanation_to_layerwise_topk_pruning} for more explanation about this pruning.


\subsection{Train Data Construction}
\label{sec:train_data_construction}
Because TreeHop optimizes next-step retrieval via contrastive learning, flat question-answer pairs are insufficient. Training requires step-by-step trajectories structured as Directed Acyclic Graphs (DAGs), whose nodes are evidence paragraphs and edges logical dependencies, exposing the model at each hop to explicit positive targets (evidence containing the bridge entity) alongside hard negatives.

To build this robust, structurally supervised dataset, we develop a three-stage pipeline designed to derive a fine-grained train dataset:

\begin{itemize}
    \item \textbf{Multi-Hop Composition:} We construct bottom-up multi-hop queries by linking single-hop questions through shared bridge entities, mapping target paragraphs to reasoning steps using LLM.
    \item \textbf{Topology and Evidence Integrity Validation:} The initial LLM-generated evidence chains can be erroneous, we validate them using another LLM, followed by strict rule-based filtering. Queries with incorrect reason topologies and those with mismatched hop-to-evidence counts are either corrected or discarded.
    \item \textbf{Contrastive Learning Graph:} Validated chains are transformed to DGL~\citep{wang2019dgl} graphs and traversed layer-by-layer in training. In each hop, current-layer passage embeddings serve as positive targets while corpus-wide passage embeddings form a negative pool. Query embeddings produced by TreeHop will be fed into TreeHop again in the next hop.
\end{itemize}

For more detailed explanation about the train data construction, please refer to \autoref{sec:detail_on_train_data}.

\subsection{Model Training} \label{sec:model_training}
We use BGE-m3~\citep{chen2024bge}, a multilingual embedding model covering over 100 languages, to embed the initial query and build the chunk embedding database, giving TreeHop potential applicability across languages\footnote{Please find our discussion about TreeHop's multilingual capability in \autoref{multilingual_capability_of_treehop}.}. BGE-m3 remains frozen during training. Prompt templates for the four datasets are in \autoref{tab:embedding_prompt_template_2wiki_musique} and \autoref{tab:embedding_prompt_template_multihop_rag} in \autoref{sec:iter_retgen_prompt_templates}.

We employ the $\mathcal{L}_{\rm{infoNCE}}$ objective in \autoref{eq:contrastive} with temperature $\tau$ of 0.15 and five negatives sampled from embedding database.
The model is compact enough to be trained on a single Nvidia V100 GPU, with batch size of 64, AdamW optimizer and learning rate of 2e-5.

\section{Experiments and Results}
To examine TreeHop, experiments are conducted regarding its retrieval performance and efficiency. Below, we introduce the selective evaluation datasets, evaluate metrics, baselines, concurrent solutions and downstream LLMs that involve in the experiments for comparison.

\begin{table*}[tbhp!]
\vspace{-10mm}

\small
\setlength{\tabcolsep}{2pt}
\begin{center}
 \setlength\tabcolsep{.75\tabcolsep}
\resizebox{\textwidth}{!}{%
\begin{tabular}{@{}lcccc|cccc|cccc|cccc@{}}
\toprule

 & \multicolumn{4}{c|}{\textbf{HOTPOT}} & \multicolumn{4}{c|}{\textbf{2WIKI}} & \multicolumn{4}{c|}{\textbf{MUSIQUE}} & \multicolumn{4}{c}{\textbf{MULTIHOP-RAG}} \\
\multicolumn{1}{l}{\textbf{Retriever}} & \multicolumn{1}{c}{\textbf{Recall}} & \multicolumn{1}{c}{\textbf{Chain}} & \multicolumn{1}{c}{\textbf{Avg.$K$}} & \multicolumn{1}{c|}{\textbf{Lat.}} & \multicolumn{1}{c}{\textbf{Recall}} & \multicolumn{1}{c}{\textbf{Chain}} & \multicolumn{1}{c}{\textbf{Avg.$K$}} & \multicolumn{1}{c|}{\textbf{Lat.}} & \multicolumn{1}{c}{\textbf{Recall}} & \multicolumn{1}{c}{\textbf{Chain}} & \multicolumn{1}{c}{\textbf{Avg.$K$}} & \multicolumn{1}{c|}{\textbf{Lat.}} & \multicolumn{1}{c}{\textbf{Recall}} & \multicolumn{1}{c}{\textbf{Chain}} & \multicolumn{1}{c}{\textbf{Avg.$K$}} & \multicolumn{1}{c}{\textbf{Lat.}}\\

\midrule

\multicolumn{16}{l}{\textit{Baselines}} \\
$\text{Direct-R@5}$ & 68.3 & 43.5 & 5 & 0.002 & 49.3 & 12.7 & 5 & 0.002 & 45.4 & 16.4 & 5 & 0.002 & 48.6 & 15.8 & 5 & 0.002 \\
$\text{Direct-R@10}$ & 71.2 & 50.5 & 10 & 0.003 & 53.2 & 16.9 & 10 & 0.003 & 49.8 & 20.2 & 10 & 0.002 & 53.0 & 19.3 & 10 & 0.002 \\
$\text{Direct-R@15}$ & 73.9 & 53.3 & 15 & 0.003 & 55.1 & 19.8 & 15 & 0.003 & 53.3 & 23.8 & 15 & 0.003 & 56.6 & 22.2 & 15 & 0.003 \\

\midrule

\multicolumn{16}{l}{\textit{Advanced RAG}} \\

$\text{\textbf{Iter-RetGen}(8B)@5 iter2}$ & 73.0 & 53.4 & 9.9 & 4.698 & 59.2 & 43.3 & 9.9 & 4.690 & \textbf{52.8} & \textbf{28.6} & 9.9 & 4.949 & 55.0 & 20.7 & 9.9 & 4.876 \\
$\text{\textbf{Iter-RetGen}(8B)@5 iter3}$ & 73.7 & 54.8 & 14.9 & 7.288 & 61.9 & 44.8 & 14.7 & 7.278 & 54.1 & \textbf{31.6} & 14.8 & 7.274 & 57.0 & 25.0 & 14.5 & 7.322 \\
$\text{\textbf{Iter-RetGen}(0.6B)@5 iter2}$ & 70.4 & 50.5 & 7.2 & 2.853 & 56.0 & 22.0 & 6.1 & 3.004 & 49.9 & 19.5 & 6.6 & 3.120 & 56.9 & 21.0 & 9.1 & 2.936 \\
$\text{\textbf{Iter-RetGen}(0.6B)@5 iter3}$ & 70.7 & 51.3 & 10.0 & 4.511 & 56.6 & 23.5 & 8.1 & 4.701 & 50.2 & 21.6 & 8.7 & 4.888 & 57.2 & 25.1 & 12.9 & 4.674 \\
$\text{\textbf{EfficientRAG}@5}$ & 71.8 & 51.8 & 4.4 & 2.851 & 60.5 & 43.8 & 3.8 & 2.846 & 46.9 & 19.1 & 6.1 & 2.907 & 51.8 & 19.0 & 4.1 & 2.855 \\
$\text{\textbf{RAPTOR}@9 collapse}$ & 71.5 & 48.6 & 9 & 5.325 & 50.9 & 15.0 & 9 & 5.483 & 49.5 & 21.2 & 9 & 5.331 & 53.6 & 20.0 & 9 & 6.250 \\
$\text{\textbf{RAPTOR}@14 collapse}$ & \textbf{75.4} & \textbf{55.1} & 14 & 5.546 & 53.3 & 17.9 & 14 & 5.709 & \textbf{54.6} & 25.7 & 14 & 5.536 & 56.0 & 20.8 & 14 & 6.108 \\

\midrule

\multicolumn{16}{l}{\textit{Ours}} \\

$\text{\textbf{TreeHop}@5 iter2}$ & \textbf{73.6} & \textbf{53.8} & 9.5 & \textbf{0.117} & \textbf{62.7} & \textbf{47.3} & 9.5 & \textbf{0.114} & 52.0 & 25.1 & 9.2 & \textbf{0.113} & \textbf{60.2} & \textbf{23.0} & 8.7 & \textbf{0.108} \\
$\text{\textbf{TreeHop}@5 iter3}$ & 74.5 & 54.9 & 14.2 & \textbf{0.202} & \textbf{66.8} & \textbf{48.1} & 14.0 & \textbf{0.204} & 53.7 & 27.7 & 13.5 & \textbf{0.200} & \textbf{66.1} & \textbf{28.9} & 12.2 & \textbf{0.181} \\

\bottomrule
\end{tabular}%
}

\caption{Results of baselines, concurrent advanced RAG solutions and TreeHop on four MHQA datasets. Direct-R@5 denotes the direct retrieval at top-5 baseline retrieval setting (See \autoref{subsec:eval_metric} for baseline and advanced RAG details). \textbf{Bold} numbers indicate the best performance among advanced retrievers at a comparable Avg.$K$.}

\label{tab:retrieve_treehop_vs_rag}

\end{center} 
\vspace{-4mm}
\end{table*}

\subsection{Datasets}
\label{sec:eval_datasets}
We benchmark TreeHop on four widely used MHQA datasets in the literature: HotpotQA~\citep{yang2018hotpotqa}, 2WikiMultiHop~\citep{xanh2020_2wikimultihop}, MuSiQue answerable~\citep{trivedi-etal-2022-musique}, and MultiHop RAG~\citep{tang2024multihopragbenchmarkingretrievalaugmentedgeneration}. Some of their questions do not challenge multi-hop retrieval performance but require LLMs to deduce from multiple documents. Since we only want to test the performance of iterative retrieval systems, we focus on question types requiring multi-step retrieval. Specifically, in 2WikiMultiHop, we filter to inference, compositional and bridge comparison questions (9,536 records), while for HotpotQA and MuSiQue, all answerable questions are used (7,405 and 2,417 records, respectively). MultiHop RAG's inference questions (816 records) are included. See \autoref{sec:dataset_cards} for detailed introduction to the types of question among the evaluate datasets,
\autoref{tab:evaluate_dataset_stats} for number of queries and size of embedding databases, and \autoref{tab:detailed_evaluate_dataset_stats} for detailed number of queries for each selective types of question.
To evaluate TreeHop's generalizability to questions that do not require multi-hop retrieval, the experimental results, \autoref{sec:experiment_all_query_types} presents results obtained using the settings described in the following subsections, but without query type filtering.

\subsection{Evaluation Metrics \& Benchmarks}
\label{subsec:eval_metric}
To evaluate retrieval performance, we report the standard recall rate under a top-$K$ retrieval setting, denoted Recall@K.
Since Recall@K is micro-averaged over gold passages, it gives partial credit to a broken chain that a downstream reader cannot use. We therefore also report \textbf{complete evidence-chain recall} (Chain), the fraction of queries whose gold passages are \textit{all} retrieved by the end of an iteration, reported cumulatively with retrieval hops. \autoref{tab:evaluate_dataset_stats} lists the average gold passages per query that this criterion demands.

To compare the efficiency among selected RAG solutions, we record the average duration per query in seconds on each dataset, denoted latency, which measures retrieval only.

\paragraph{End-to-End QA Evaluation}
Following Efficient RAG and Iter-Retgen, we employ exact match (EM), as well as accuracy (ACC), a metric that evaluates model answers by GPT-5-mini, to evaulate end-to-end QA performance among iterative retrieval solutions. The prompt can be found in \autoref{sec:acc_prompt_gpt5mini}.
We adopt Llama3.1-8B-Instruct~\citep{grattafiori2024llama3herdmodels}, Qwen3.6-27B~\citep{qwen3.6-27b} with reasoning capability, and Qwen2.5-7B-Instruct~\citep{qwen2025qwen25technicalreport} as downstream LLMs. Decoding hyper-parameters and the QA prompt adopted for the three models are in \autoref{sec:qa_prompt_for_e2e_eval}.

\paragraph{Baselines and Advanced RAG} Our baseline is native top-$K$ retrieval, and all methods use BGE-m3 as the retrieval backbone for fairness. To isolate the benefit of structural optimization, we confine the comparison to TreeHop's own paradigm, efficient iterative retrieval, benchmarking against \mbox{Iter-RetGen}~\citep{shao2023enhancingretrievalaugmentedlargelanguage} and EfficientRAG~\citep{zhuang2024efficientragefficientretrievermultihop} with Filter (0.3B) and Labeler (0.4B) models. We refrain from comparing against heavy LLM frameworks, whose parametric reasoning capacity would confound the comparison. We additionally compare against RAPTOR~\citep{sarthi2024raptor}, which is similar in the tree structured reasoning graph. It offers two retrieval modes, of which we report collapse mode that performs better; \autoref{sec:raptor_comparison} gives full mode details.

We instantiate Iter-RetGen at two backbone scales, Llama3.1-8B-Instruct~\citep{llama3modelcard} and the compact Qwen3-0.6B~\citep{yang2025qwen3technicalreport}, and evaluate both Iter-RetGen and TreeHop at their second and third retrieval iterations to expose behaviour at different depths. Prompt templates for Iter-RetGen are in \autoref{tab:iter-retgen_prompt_template}.

\subsection{Results}
\label{sec:results}
We report results of TreeHop and benchmarks on four datasets; \autoref{sec:case_study_treehop_retrieval} contains a case study on the retrieval outcomes.

\paragraph{Retrieval Efficiency}
As shown in \autoref{tab:retrieve_treehop_vs_rag}, TreeHop reduces computational overhead by more than an order of magnitude while remaining competitive on recall, at 0.1 seconds per query in the second iteration and 0.2 seconds in the third, over 2.6 seconds faster than the next best solution, EfficientRAG. This is a latency reduction of 95.9\%-96.2\% at the second iteration and 92.8\%-93.7\% at the third, of 97.2\%-97.8\% against Iter-RetGen(8B), and of 96.4\%-98.3\% against RAPTOR. It stems from TreeHop's embedding-level computation, which avoids the recursive token generation loops of LLM-based methods.

\begin{table*}[t!]
\vspace{-10mm}
\footnotesize
\renewcommand{\arraystretch}{1.}
\begin{center}
 \setlength\tabcolsep{.75\tabcolsep}
\begin{tabular}{@{}ll|cc|cc|cc|cc@{}}
\toprule
& & \multicolumn{2}{c|}{\textbf{HOTPOT}} & \multicolumn{2}{c|}{\textbf{2WIKI}} & \multicolumn{2}{c|}{\textbf{MUSIQUE}} & \multicolumn{2}{c}{\textbf{MULTIHOP-RAG}} \\
\multicolumn{1}{c|}{\textbf{Method}} & \multicolumn{1}{c|}{\textbf{Downstream LM}} & \textbf{EM} & \textbf{ACC} & \textbf{EM} & \textbf{ACC} & \textbf{EM} & \textbf{ACC} & \textbf{EM} & \textbf{ACC} \\
\midrule
\multirow{2}{*}{$\text{\textbf{Iter-RetGen}(8B)@5 iter2}$}
 & Qwen 3.6 & 62.4 & 66.0 & 52.1 & 54.8 & \textbf{46.4} & \textbf{50.0} & 98.7 & 99.0 \\
 & Qwen 2.5 & 43.6 & 46.5 & 38.0 & 40.6 & \textbf{25.4} & \textbf{27.0} & 89.8 & \textbf{91.1} \\
 & Llama 3.1   & \textbf{34.2} & \textbf{37.0} & 30.5 & 32.3 & \textbf{19.1} & 19.9 & 86.3 & 87.8 \\
\addlinespace[3.5pt]
\multirow{2}{*}{$\text{\textbf{Iter-RetGen}(8B)@5 iter3}$}
 & Qwen 3.6 & 63.0 & 66.8 & 53.3 & 56.1 & \textbf{49.3} & \textbf{52.3} & 98.9 & 99.1 \\
 & Qwen 2.5 & 43.9 & 46.7 & 39.8 & 41.7 & \textbf{27.8} & \textbf{29.7} & 84.5 & 86.5 \\
 & Llama 3.1   & 30.9 & 33.7 & 26.2 & 28.5 & \textbf{18.0} & \textbf{21.0} & 83.9 & 86.4 \\
\addlinespace[3.5pt]
\multirow{2}{*}{$\text{\textbf{Iter-RetGen}(0.6B)@5 iter2}$}
 & Qwen 3.6 & 58.2 & 61.3 & 48.5 & 51.3 & 23.5 & 26.2 & 98.8 & 99.0 \\
 & Qwen 2.5 & 42.2 & 44.4 & 34.4 & 37.3 & 20.5 & 23.1 & 89.9 & 90.6 \\
 & Llama 3.1   & 33.0 & 34.8 & 27.4 & 30.1 & 14.3 & 17.0 & 86.3 & 87.4 \\
\addlinespace[3.5pt]
\multirow{2}{*}{$\text{\textbf{Iter-RetGen}(0.6B)@5 iter3}$}
 & Qwen 3.6 & 58.2 & 61.6 & 49.2 & 52.1 & 24.5 & 27.7 & 98.8 & 99.0 \\
 & Qwen 2.5 & 43.0 & 45.3 & 35.3 & 37.9 & 21.2 & 23.7 & 85.6 & 87.6 \\
 & Llama 3.1   & 28.6 & 31.4 & 23.0 & 25.4 & 14.0 & 16.4 & 84.3 & 86.7 \\
\addlinespace[3.5pt]
\multirow{2}{*}{$\text{\textbf{EfficientRAG}@5}$}
 & Qwen 3.6 & 59.8 & 62.1 & 52.7 & 55.3 & 21.2 & 22.6 & 96.8 & 97.4 \\
 & Qwen 2.5 & 42.8 & 44.6 & 38.7 & 41.0 & 18.2 & 19.5 & 88.7 & 89.5 \\
 & Llama 3.1   & 33.4 & 35.7 & 31.2 & 33.4 & 15.9 & 17.8 & 86.0 & 86.9 \\
\addlinespace[3.5pt]
\multirow{2}{*}{$\text{\textbf{RAPTOR}@9 collapse}$}
 & Qwen 3.6 & 57.9 & 60.6 & 47.7 & 49.7 & 35.2 & 38.7 & 97.3 & 98.2 \\
 & Qwen 2.5 & 40.4 & 42.5 & 32.6 & 34.8 & 21.0 & 23.4 & 89.1 & 90.4 \\
 & Llama 3.1   & 33.4 & 35.0 & 21.4 & 23.5 & 18.8 & \textbf{20.9} & \textbf{87.5} & 87.9 \\
\addlinespace[3.5pt]
\multirow{2}{*}{$\text{\textbf{RAPTOR}@14 collapse}$}
 & Qwen 3.6 & \textbf{65.1} & \textbf{68.2} & 48.1 & 50.6 & 41.3 & 44.6 & 98.7 & 99.0 \\
 & Qwen 2.5 & \textbf{45.3} & \textbf{48.1} & 33.9 & 35.3 & 23.1 & 25.3 & 89.7 & 90.8 \\
 & Llama 3.1   & 30.3 & 34.0 & 20.3 & 23.2 & 17.6 & 19.4 & \textbf{86.5} & \textbf{87.7} \\
\midrule
\multirow{2}{*}{$\text{\textbf{TreeHop}@5 iter2}$}
 & Qwen 3.6 & \textbf{62.7} & \textbf{66.5} & \textbf{56.3} & \textbf{58.9} & 43.8 & 46.5 & \textbf{99.0} & \textbf{99.2} \\
\textbf{} & Qwen 2.5 & \textbf{43.8} & \textbf{46.7} & \textbf{41.6} & \textbf{43.1} & 24.0 & 26.4 & \textbf{90.0} & 90.8 \\
 & Llama 3.1   & 34.0 & 36.2 & \textbf{32.3} & \textbf{34.5} & 18.8 & 19.8 & 87.0 & \textbf{88.2} \\
\addlinespace[3.5pt]
\multirow{2}{*}{$\text{\textbf{TreeHop}@5 iter3}$}
 & Qwen 3.6 & 64.4 & 67.8 & \textbf{61.9} & \textbf{64.1} & 46.9 & 50.8 & \textbf{99.0} & \textbf{99.3} \\
 & Qwen 2.5 & 44.7 & 47.7 & \textbf{45.8} & \textbf{48.3} & 26.0 & 28.9 & \textbf{91.4} & \textbf{92.1} \\
 & Llama 3.1   & 32.4 & 34.5 & \textbf{32.2} & \textbf{35.1} & 17.2 & 18.8 & 86.4 & 87.3 \\
\bottomrule
\end{tabular}
\caption{End-to-End Question Answering (QA) Results. \textbf{Bold} numbers indicate the best model answer performance among advanced iterative retrieval frameworks in a same retrieval iteration step.}
\label{tab:e2e_treehop_vs_rag}
\end{center}
\vspace{-4mm}
\end{table*}

\paragraph{Retrieval Performance}
TreeHop achieves strong performance across datasets while balancing efficiency and effectiveness. On the HotpotQA dataset, TreeHop’s recall leads all iterative solutions at the second retrieval iteration and trails only RAPTOR at the larger budget of the third;
On the 2WikiMultiHop and Multihop dataset, TreeHop surpasses Iter-RetGen backboned by Llama3.1-8B-Instruct by 3.5\%-5.2\% recall in the second iteration and 4.9\%-9.1\% recall in the third iteration, with slightly less chunks retrieved on average. This demonstrates the effectiveness of the embedding mechanism in \autoref{eq:treehop_architecture2}.
For the MuSiQue dataset, recall is 0.4\% lower than Iter-RetGen, likely due to the dataset's requirement for synthesizing information from multiple chunks (e.g., branching and converging paths in \autoref{tab:dataset_cards2}): TreeHop predicts embeddings from query-chunk pairs, which limits synthesis across several chunks at once. EfficientRAG degrades on this dataset as well, suggesting a limitation common to the query-chunk-pair strategy.

\paragraph{Complete Evidence-Chain Recall}
Direct retrieval closes the full evidence chain for only 12.7\%-43.5\% of queries, far below its 45.4\%-68.3\% micro recall. Iteration recovers that gap: by the third iteration TreeHop raises Chain by 11.3\%-35.4\% over direct retrieval, exceeding the corresponding micro recall gain on three of four datasets, and leads every concurrent solution on 2WikiMultiHop and MultiHop-RAG. Iter-RetGen(8B) retains its MuSiQue advantage (31.6\% against 27.7\%) for the synthesis reason above, and RAPTOR edges ahead on HotpotQA (55.1\% against 54.9\%) at a comparable Avg.$K$. Iterating also buys more chains than simply widening the retrieval window. At a comparable Avg.$K$, TreeHop@5 iter3 gains 1.6\%-28.3\% Chain over Direct-R@15 on all four datasets, yet its recall gain there is only 0.6\% on HotpotQA and 0.4\% on MuSiQue: retrieving more documents recovers isolated gold passages about as well as iterating, but it is completing the chain that iterative rewriting contributes.

\paragraph{Average Number of $K$}
The average number of retrieved chunks grows linearly with retrieval step, a predictability that serves as a practical deployment guideline. Our stopping criteria (see~\autoref{sec:stop_criterion}) achieve this by reducing the theoretical exponential growth under top-5 retrieval from $5^2=25$ chunks to 8.7–9.5 in the second iteration and from $5^3=125$ to 12.2–14.2 in the third.

\paragraph{End-to-End QA performance}
As illustrated in ~\autoref{tab:e2e_treehop_vs_rag}, downstream LMs achieve the best end-to-end performance on 2WikiMultiHop and MultiHop-RAG with TreeHop’s retrieval outcomes, while on MusiQue they work the best with Iter-Retgen’s retrieval results and on HotpotQA with RAPTOR’s at $K=14$. This performance is correlated to the performance of upstream retrieval systems, where TreeHop also leads in chain on 2WikiMultiHop and MultiHop-RAG, rivals Iter-Retgen on HotpotQA and underperforms Iter-Retgen on MusiQue, as shown in ~\autoref{tab:retrieve_treehop_vs_rag}. What the third iteration is worth to a reader depends on how much context it can absorb. Llama3.1-8B-Instruct, whose context window is the shortest of the three, loses exact match on all four datasets as average $K$ grows from the second iteration to the third, whereas both Qwen models improve on every one of them. The stopping criterion is what makes this manageable: by holding average $K$ to 12.2--14.2 rather than the $5^3$ of unpruned expansion, it keeps the retrieved context within reach of a short-context reader.


\section{Ablation Study}

\subsection{Effectiveness of Architecture} \label{subsec:ablation_architecture}
To evaluate the necessity of each component in \autoref{eq:treehop_architecture1}, we ablated the query term $q$, the $ForgetGate$, and the $UpdateGate$ in isolation. Each variant was trained 10 times with identical hyperparameters (as in \autoref{sec:model_training}), and performance was evaluated on the second hop. Results are shown in \autoref{tab:ablation_TreeHop_architecture} and analyzed below.

Removing $ForgetGate$ has the least impact, costing 3.3\%-4.6\% average recall: $UpdateGate$ still retains the critical chunk information. Training convergence time nonetheless rose by roughly 15\%, as the model struggled to suppress redundant information without the subtractive branch.

Without $q$, the model loses critical information from query, thereby experiences significant performance degrade on all the datasets, improves only 0.9\%-6.4\% over the vanilla top-5 retrieval.

Without $UpdateGate$, recall dropped back to the direct retrieval baseline, matching Direct-R@5 on all four datasets, confirming the gate’s critical role in integrating new information. Without it, the model degenerated to a gated subtraction that can only remove information from the query, failing to refine query embeddings iteratively.


\begin{table}[tbp]
\vspace{-10mm}
\centering
\scriptsize
\setlength{\tabcolsep}{1.5pt}
\resizebox{\columnwidth}{!}{%
\begin{tabular}{lcccc}
\toprule
\multicolumn{1}{l}{\textbf{Architecture}} & \multicolumn{1}{c}{\textbf{HOTPOT}} & \multicolumn{1}{c}{\textbf{2WIKI}} & \multicolumn{1}{c}{\textbf{MUSIQUE}} & \multicolumn{1}{c}{\makecell{\textbf{MULTIHOP}\\\textbf{RAG}}} \\
\midrule
\multicolumn{1}{l}{Direct-R@5} & 68.3 & 49.3 & 45.4 & 48.6 \\
\multicolumn{1}{l}{\textbf{TreeHop}@5 iter2} & 73.6 & 62.7 & 52.0 & 60.2 \\

\midrule

\textit{w/o. $ForgetGate$} & 70.3 (3.3\textit{$\downarrow$}) & 58.6 (4.1\textit{$\downarrow$}) & 48.1 (3.9\textit{$\downarrow$}) & 55.6 (4.6\textit{$\downarrow$}) \\

\textit{w/o. $q$} & 69.2 (4.4\textit{$\downarrow$}) & 55.7 (7.0\textit{$\downarrow$}) & 46.6 (5.4\textit{$\downarrow$}) & 51.4 (8.8\textit{$\downarrow$}) \\

\textit{w/o. $UpdateGate$} & 68.3 (5.3\textit{$\downarrow$}) & 49.3 (13.4\textit{$\downarrow$}) & 45.4 (6.6\textit{$\downarrow$}) & 48.6 (11.6\textit{$\downarrow$}) \\

\bottomrule
\end{tabular}%
}

\caption{Ablation study on TreeHop model architecture. The TreeHop experiences degraded performances when core components $q$, $ForgetGate$ and $UpdateGate$ are removed from its architecture, demonstrating their functionalities to the model performance. Recall is averaged over the 10 training runs described in \autoref{subsec:ablation_architecture}.}

\label{tab:ablation_TreeHop_architecture}

\vspace{-4mm}
\end{table}

\subsection{Effectiveness of Stopping Criterion}
The stopping criterion serves for filtering query paths to reduce computational cost without sacrificing too much performance.
In \autoref{tab:ablation_stop_criterion}, we examine the performance without the presence of Redundancy Pruning and Layer-wise Top Pruning.

\paragraph{Redundancy Pruning}
Redundancy pruning prevents revisiting previously retrieved chunks. \autoref{tab:ablation_stop_criterion} shows that removing this pruning increases the average number of retrieved chunks $K$ to 10, but reduces recall by 0.5\%-4.8\%. This occurs because when cooperate with layer-wise top pruning, redundancy pruning further ensures only unique, informative paths are pursued, thereby maintaining recall performance. Without redundancy pruning, more duplicated information take the place, resulting in degraded performance.

\paragraph{Layer-wise Top Pruning}
This pruning strategy selects top-5 chunks at each layer to control branching. Removing this strategy increases the average number of retrieved chunks by 63\%-94\% across datasets, but yields a 3.6\%-14.5\% improvement in recall. This suggests that the pruning introduces a trade-off between computational efficiency and recall performance.

\begin{table}[tbp]
\vspace{-10mm}
\centering
\scriptsize
\setlength{\tabcolsep}{1.5pt}
\resizebox{\columnwidth}{!}{%
\begin{tabular}{lcccccccc}
\toprule
 & \multicolumn{2}{c}{\textbf{HOTPOT}}& \multicolumn{2}{c}{\textbf{2WIKI}} & \multicolumn{2}{c}{\textbf{MUSIQUE}} & \multicolumn{2}{c}{\makecell{\textbf{MULTIHOP}\\\textbf{RAG}}} \\
 \cline{2-9}\\
\multicolumn{1}{l}{\textbf{Stopping Criterion}} & \multicolumn{1}{c}{\textbf{Recall}} & \multicolumn{1}{c}{\textbf{Avg.$K$}} &
\multicolumn{1}{c}{\textbf{Recall}} & \multicolumn{1}{c}{\textbf{Avg.$K$}} &
\multicolumn{1}{c}{\textbf{Recall}} & \multicolumn{1}{c}{\textbf{Avg.$K$}} &
\multicolumn{1}{c}{\textbf{Recall}} & \multicolumn{1}{c}{\textbf{Avg.$K$}} \\
\midrule
\multicolumn{1}{l}{\textbf{TreeHop}@5 iter2} & 73.6 & 9.5 & 62.7 & 9.5 & 52.0 & 9.2 & 60.2 & 8.7 \\

\midrule

\makecell[l]{\textit{w/o. Redundancy}\\\textit{Pruning}} & 70.4 & 10 & 59.9 & 10 & 47.2 & 10 & 59.7 & 10 \\
\makecell[l]{\textit{w/o. Layer-wise}\\\textit{Top Pruning}} & 79.1  & 17.0 & 77.2  & 18.4 & 55.6 & 16.6 & 69.3 & 14.2 \\
\textit{w/o. Both} & 79.1 & 30 & 77.2 & 30 & 55.6 & 30 & 69.4 & 30 \\

\bottomrule
\end{tabular}%
}

\caption{Ablation study on stopping criterion, where redundancy pruning and layer-wise top 5 pruning are removed from post retrieval process, respectively.
}

\label{tab:ablation_stop_criterion}

\vspace{-4mm}
\end{table}

\paragraph{Combined Effects}
Combining redundancy pruning and layer-wise top pruning trades recall for a substantially smaller retrieval budget. Take Multihop RAG for example, when both criteria are applied, TreeHop achieves a recall of 60.2\% with an average number of retrieved chunks 8.7; disabling both raises recall to 69.4\%, but at 30 chunks in the second iteration, a 3.4$\times$ increase. Normalising recall by the number of retrieved chunks makes the trade explicit: 6.9 recall per chunk with pruning against 2.3 without, a 3.0$\times$ gain in retrieval efficiency for a 9.2\% recall cost. Layer-wise top pruning limits the branching factor while redundancy pruning keeps the retained paths distinct, together averting the exponential path explosion inherent to iterative systems.

\section{Conclusion}
We introduce TreeHop, a lightweight query embedding model that refines query embeddings across retrieval iterations without invoking an additional LLM or a rewrite component. Its core mechanism is a pair of embedding-level cross-attention gates: the $UpdateGate$ selectively integrates information from retrieved chunks, and the $ForgetGate$ discards the query semantics those chunks have already resolved. Paired with a simple rule-based stopping criterion, this keeps the model at 0.18 billion parameters while remaining competitive on four MHQA benchmarks.
Future work could explore more effective model architectures, adaptive stopping criteria, or extensions to lengthy, structural, or multi-modal inputs. Embedding-centric rewriting offers a practical balance between retrieval performance and computational cost, paving the way for real-world multi-hop reasoning in industrial applications.

\section{Limitation} \label{sec:limitation}
Notwithstanding our method have demonstrated efficacy,
our TreeHop's query-rewriting mechanism depends on query-chunk pair embeddings, limiting its capability to synthesize information across multiple chunks simultaneously. This design choice prioritizes efficiency and system complexity, as synthesizing arbitrary numbers of chunks requires additional tools or models to repel irrelevant chunks that can potentially contaminate the produced query embeddings.
We also note that our TreeHop has the potential to further improve its performance, please find our analysis on the scale of train dataset in \autoref{sec:analysis_on_the_size_of_training_dataset}.

\clearpage

\bibliography{references}

\clearpage

\appendix

\clearpage

%
%
\section{Dataset Cards} \label{sec:dataset_cards}
Below illustrates datasets inclusive in our work, the question types for evaluation are selected to ensure synthesizing information from query and retrieved document chunks are mandatory for multihop retriever.

%
\begin{table}[h!]
\hsize\textwidth \linewidth\hsize
\centering
\small
\renewcommand\tabcolsep{0.8pt} 
\begin{tabular}{cp{9cm}cp{2cm}}
    \toprule
    \textbf{Dataset} & \textbf{Question Type} & \textbf{Require Synthesize} \\
    \midrule
    \multirow{22}{*}{\parbox{2.5cm}{\centering 2WikiMultiHop}}

    & \textbf{Comparison question}: Questions requiring direct comparison of attributes between entities within the same category.\\
    & \textit{Example Question}: Who was born first, Albert Einstein or Abraham Lincoln? \\

    \cmidrule(lr){2-3}

    & \textbf{Inference question}: Questions requiring derivation of implicit relationships by combining triples from a knowledge graph.\\
    & \textit{Example Question}: Who is the maternal grandfather of Abraham Lincoln?  & \huge \checkmark \\
    & \textit{Triples}: (Abraham Lincoln, mother, Nancy Hanks Lincoln); (Nancy Hanks Lincoln, father, James Hanks). \\

    \cmidrule(lr){2-3}

    & \textbf{Compositional question}: Questions requiring multi-step relational reasoning across non-explicitly linked triples.\\
    & \textit{Example Question}: Who founded the distributor of \textit{La La Land}?  & \huge \checkmark \\
    & \textit{Triples}: (La La Land, distributor, Summit Entertainment); (Summit Entertainment, founded by, Bernd Eichinger). \\

    \cmidrule(lr){2-3}

    & \textbf{Bridge-comparison question}: Questions requiring both bridging to intermediate entities and comparative reasoning.\\
    & \textit{Example Question}: Which movie has the director born first, \textit{La La Land} or \textit{Tenet}?  & \huge \checkmark \\
    & \textit{Steps}: 1. Find directors: \textit{La La Land} $\rightarrow$ Damien Chazelle; \textit{Tenet} $\rightarrow$ Christopher Nolan. \\
    & 2. Compare birth years: Damien Chazelle (1985) vs. Christopher Nolan (1970). \\

    \midrule

    \multirow{16}{*}{\parbox{2cm}{\centering MuSiQue}} &

    \textbf{Unanswerable}: Questions with potential support paragraphs are partially removed, making the reasoning infeasible or unable to arrive at the correct answer. \\

    \cmidrule(lr){2-3}

    & \textbf{3-Hop Reasoning (Linear Path)} A sequential, three-step logical connection. \\
    & \textit{Example Question}: What currency is used where Billy Giles died?  & \huge \checkmark \\
    & \textit{steps}: 1. Find the location of Billy Giles' death. \\
    & 2. Locate the region this place belongs to. \\
    & 3. Identify the currency used in that region. \\

    \cmidrule(lr){2-3}

    & \textbf{3-Hop Reasoning (Branching-Converging Path)} Begins with a single inquiry but diverges into different, branching sub-questions, then converges. \\
    & \textit{Example Question}: When was the first establishment that McDonaldization is named after, opened in the country Horndean is located?  & \huge \checkmark \\
    & \textit{steps}: 1. Determine what McDonaldization refers to. \\
    & 2. Identify the country where Horndean is located. \\
    & 3. Find the date the first establishment opened in that country. \\

\end{tabular}
\caption{Dataset Cards.}
\label{tab:dataset_cards1}
\end{table}

\clearpage

\begin{table}[h!]
\hsize\textwidth \linewidth\hsize
\centering
\small
\renewcommand\tabcolsep{0.8pt} 
\begin{tabular}{cp{9cm}cp{2cm}}
    \midrule
    \textbf{Dataset} & \textbf{Question Type} & \textbf{Require Synthesize} \\
    \midrule

    \multirow{48}{*}{\parbox{2cm}{\centering MuSiQue}} &
    \textbf{2-Hop Reasoning (Linear Path)}: A single, straightforward logical path connecting two facts. \\
    & \textit{Example Question}: Who succeeded the first President of Namibia?  & \huge \checkmark \\
    & \textit{steps}: 1. Identify the first President of Namibia. \\
    & 2. Determine who succeeded them. \\

    \cmidrule(lr){2-3}

    & \textbf{4-Hop Reasoning (Linear Path)} A continuous, four-step logical progression. \\
    & \textit{Example Question}: When did Napoleon occupy the city where the mother of the woman who brought Louis XVI style to the court died? & \huge \checkmark \\
    & \textit{steps}: 1. Identify who introduced Louis XVI style. \\
    & 2. Find their mother. \\
    & 3. Determine the city of the mother’s death. \\
    & 4. Discover when Napoleon occupied that city. \\

    \pagebreak \\
    \cmidrule(lr){2-3}

    & \textbf{4-Hop Reasoning (Branching-Converging Path)} Starts with a single query, splits into multiple paths, and then converges. \\
    & \textit{Example Question}: How many Germans live in the colonial holding in Aruba’s continent that was governed by Prazeres’s country? & \huge \checkmark \\
    & \textit{steps}: 1. Locate Aruba's continent. \\
    & 2. Identify Prazeres’ country. \\
    & 3. Determine the colonial holding governed by that country in Aruba’s continent. \\
    & 4. Find the number of Germans there. \\

    \cmidrule(lr){2-3}

    & \textbf{4-Hop Reasoning (Converging Path)}: Multiple distinct lines of reasoning that eventually converge on the answer. \\
    & \textit{Example Question}: When did the people who captured Malakoff come to the region where Philipsburg is located? & \huge \checkmark \\
    & \textit{steps}: 1. Determine Philipsburg's location. \\
    & 2. Identify the terrain feature it belongs to. \\
    & 3. Find who captured Malakoff. \\
    & 4. Determine when those people came to that terrain. \\

    \midrule

    \multirow{8}{*}{\parbox{2cm}{\centering MultiHop RAG}} & 

    \textbf{Inference Query}: Questions requiring derivation of implicit relationships by combining triples from a knowledge graph.\\
    & \textit{Example Question}: Who is the maternal grandfather of Abraham Lincoln?  & \huge \checkmark \\
    & \textit{Triples}: (Abraham Lincoln, mother, Nancy Hanks Lincoln); (Nancy Hanks Lincoln, father, James Hanks). \\

    \cmidrule(lr){2-3}

    & \textbf{Comparison query}: Questions requiring direct comparison of attributes between entities within the same category.\\
    & \textit{Example Question}: Did Netflix or Google report higher revenue for the year 2023? \\

    \cmidrule(lr){2-3}

    & \textbf{Temporal query}: Question that
requires an analysis of the temporal information
of the retrieved chunks \\
    & \textit{Example Question}: Did Apple introduce the AirTag
tracking device before or after the launch of the 5th generation iPad Pro? \\

    \cmidrule(lr){2-3}

    & \textbf{Null query}: Question whose answer cannot
be derived from the retrieved set.This is purposely for testing the issue of hallucination. The LLM should produce a null response instead of hallucinating an answer. \\
    & \textit{Example Question}: What are the sales of company ABCD as reported in its 2022 and 2023 annual reports? \\

    \bottomrule
\end{tabular}
\caption{Dataset Cards.}
\label{tab:dataset_cards2}
\end{table}

\clearpage

\section{Details on Evaluate Datasets}
We illustrate number of records in all inclusive datasets in the following table.

\begin{table}[ht!]
\hsize\textwidth \linewidth\hsize
    \centering
    \begin{tabular}{cccc}
    \toprule
        Dataset & Query & Embedding Database & Avg. Gold Passages \\
    \midrule
        HotpotQA-dev-distractor & 7,405 & 66,635 & 2.00 \\
        2WikiMultihop-dev & 9,536 & 56,709 & 2.58 \\
        MuSiQue-dev & 2,417 & 21,100 & 2.60 \\
        Multihop-RAG-dev & 816 & 609 & 2.47 \\
    \bottomrule
    \end{tabular}
    \caption{Descriptive statistics of datasets in terms of the number of queries, sizes of the embedding database, and the average number of gold supporting passages per query, which is the number of passages a retriever must recover to close the evidence chain.}

    \label{tab:evaluate_dataset_stats}

\end{table}

Below, we provide detailed number of questions for each question type in our evaluate datasets. Please refer to \autoref{sec:dataset_cards} for introduction to these types.
\begin{table}[ht!]
\hsize\textwidth \linewidth\hsize
    \centering
    \begin{tabular}{ccc}
    \toprule
    \textbf{Dataset} & \textbf{Question Type} & \textbf{Count} \\
    \midrule
    \multirow{1}{*}{\parbox{2.5cm}{\centering HotpotQA}}

    & - & 7,405 \\

    \midrule
    \multirow{3}{*}{\parbox{2.5cm}{\centering 2WikiMultiHop}}

    & Compositional & 5,236 \\
    & Bridge Comparison & 2,751 \\
    & Inference & 1,549 \\

    \midrule
    \multirow{6}{*}{\parbox{2.5cm}{\centering MuSiQue}}
    & 2-hop reasoning (linear path) & 1,252 \\
    & 3-Hop Reasoning (Linear Path) & 568 \\
    & 3-Hop Reasoning (Branching Path) & 192 \\
    & 4-Hop Reasoning (Linear Path) & 246 \\
    & 4-Hop Reasoning (Branching Path) & 64 \\
    & 4-Hop Reasoning (Converging Path) & 95 \\

    \midrule

    \multirow{1}{*}{\parbox{2.5cm}{\centering Multihop RAG}}

    & Inference & 816 \\
    \bottomrule
    \end{tabular}
    \caption{Evaluate data statistics on number of queries and sizes of embedding database.}
    \label{tab:detailed_evaluate_dataset_stats}
\end{table}

\clearpage

\section{Details on Train Data}
\label{sec:detail_on_train_data}
Due to the space limitations, we refrained from introducing low‑level construction steps on our train data in the main text; however, this process is non-trivial.
The TreeHop model fundamentally uses contrastive learning to learn next query embedding given a query-chunk embedding pair. Therefore, training TreeHop requires embedding level dataset structured using multi-hop reasoning steps, with each step provides the model with positive-negative passage pairs, where positive passages contain evidence steer toward ground truths. This results in directed acyclic graphs (DAGs) that mirrors the reasoning path. Below, we describes the sources of initial questions, the derivation of directed evidence chains, the pipeline for turning raw data into DGL training graphs, the augmentation strategy for comparison queries, and cleaning stages.

\subsection{From Single-hop Question to Multi-hop Question}
We started with collecting single-hop questions from four English reading comprehension datasets: SQuAD~\citep{rajpurkar2016squad100000questionsmachine}, Natural Questions~\citep{kwiatkowski-etal-2019-natural}, MLQA (en-en)~\citep{munis2025fibermultilingualevaluationresource} and T-REx~\citep{elsahar2018t}. Multi-hop questions can be viewed as a series of single-hop sub-questions and are constructed bottom-up from a sequence of corresponding reasoning steps, where answer from one reasoning step was used to identify the next reasoning step. Following ~\citep{trivedi-etal-2022-musique}, we adopt the same methodology to derive multi-hop questions, with each multihop question $Q$ has an underlying DAG $G_Q$ whose nodes are single-hop sub-questions $q_1,q_2,\ldots,q_n$ and where a directed edge $(q_j,q_i)$ indicates that $q_i$ depends on the answer of $q_j$ via a shared bridge entity~\citep{trivedi-etal-2022-musique}. For example, the multi-hop question \textit{"Who succeeded the first President of
Namibia?"}, is composited by two single-hop sub-questions: \textit{"Who was the first President of Namibia?"} and "Who succeeded ..." where the answer of the first question is the bridge entity, we annotate paragraphs in which the bridge entities reside as the supporting paragraphs.

\subsection{Evidence Chain Validation}
\label{appendix:evidence_validation}
After generating questions and their supporting paragraphs, we employ LLM-based inference to validate the correct reasoning order among the supporting paragraphs. This is crucial because they were generated by LLMs and are not 100\% reliable.
Concretely, for each sample we extract the graph type from its sample ID, look up the expected reasoning topology, and construct a structured prompt containing the question, its answer, and the set of supporting paragraphs in random order. A reasoning model is then prompted to output a JSON list of \texttt{\{title, next\}} pairs specifying how the supporting paragraphs chain together. Structured output constraints ensure the model produces valid JSON conforming to the expected schema, and a maximum of 3 retries are scheduled to recover from parsing failures.

The LLM responses are post-processed with fuzzy title matching to resolve minor discrepancies between the model's output and the actual paragraph titles. Each parsed chain is then converted into evidence triplets $(p_j, \cdot, p_i)$---where paragraph $p_j$ leads to paragraph $p_i$---and validated: we verify that the resulting directed graph is acyclic, has the correct number of nodes for the expected topology, and that every source title appears in the sample's context paragraphs. Samples that fail validation are retried with increased temperature; those that still fail after all retries are discarded.

The prompt template for validating the evidence chain, and a corresponding instantiated example can be found in~\autoref{tab:prompt_template_for_evidence_chain_validation} and~\autoref{tab:evidence_chain_template_example}, respectively.

\subsection{Vectorization}
We employ a multilingual embedding model, BGE-m3~\citep{chen2024bge}, to vectorize queries and document chunks. The BGE-m3 is frozen throughout our training, isolating the unique contribution of performance improvement by our framework. We adopt Faiss~\citep{johnson2019billion} IndexFlatIP for vector database construction and retrieval similarity searching.

\subsection{Graph Construction Pipeline}

For each training sample we parse its annotated evidence chain into a directed graph $G=(V,E)$ whose nodes are passages and whose edges encode hop-level dependencies. We verify that $G$ is acyclic and that its topology matches the expected question type (e.g., a single entry node for compositional questions, parallel entry nodes for comparison questions); samples that fail these checks are discarded. We then traverse $G$ in breadth-first order: at each layer the current-layer passages serve as positive next-hop targets, while a set of embeddings randomly drawn from the corpus pool (excluding the current and future positive nodes) serve as negatives. This layer-wise decomposition yields one contrastive training group per hop, in direct correspondence with the hop-indexed retrieval at inference time.

On a higher level, this graph formulation mimics the DAG, i.e., the inference-time hop sequence, enabling contrastive training with structural and scalable supervision.

\subsection{Comparison Question Augmentation}
\label{appendix:comparison_augmentation}
Comparison questions are questions that compare two or more entities (see~\autoref{sec:dataset_cards}). More generally, for a comparison question with two first-layer passages $P_a$ and $P_b$, we generate two training graphs:
\begin{enumerate}
    \item $\text{query} \to P_a \to P_b$\,: first retrieves $P_a$, then (conditioned on the rewritten query) retrieves $P_b$ as a leaf node.
    \item $\text{query} \to P_b \to P_a$\,: the retrieval order is reversed.
\end{enumerate}
Each variant is an independent training graph with its own negative samples drawn per hop. This augmentation doubles the number of training examples originating from comparison questions and ensures that TreeHop learns to sequentially retrieve both passages regardless of the order in which they appear.

We summarize the goals and the remaining data record for each stages in the following table.

%
%
\begin{table}[ht!]
    \centering
    \setlength{\tabcolsep}{3pt}
    \begin{tabular}{p{6.0cm}p{7.6cm}r}
    \toprule

    \textbf{Cleaning Stage} & \textbf{Goal} & \makecell[r]{\textbf{Remaining}\\\textbf{Records}} \\

    \midrule

    \textit{Initial Train Data} & & 233,487 \\

    \midrule

    \textbf{Topology Validation} & Ensure the presence (need) of a follow‑up retrieval step and the graph is acyclic. & 215,378 \\

    \midrule

    \textbf{Evidence Integrity Check} & Verify that the generated reasoning topology matches the evidence chain. (e.g., a “three‑hop” question can only have exactly two follow‑up queries) & 178,394 \\

    \midrule

    \textbf{Comparison Question Augmentation} & Enrich trainable data & 249,133 \\

    \bottomrule
    \end{tabular}
\end{table}

\clearpage

\section{Prompt Template for Evidence Chain Validation}
\label{sec:prompt_templates_for_evidence_chain_validation}
We present prompt templates for validating evidence chain. This ensures the evidence chain matches the question composition path, which is crucial for a fine-grained train dataset.

\begin{table}[ht!]
\hsize\textwidth \linewidth\hsize
    \centering
    \begin{tabular}{p{15cm}}
    \toprule
        \textbf{Prompt Template for Evidence Chain Validation} \\
    \midrule
\textbf{\textit{SYSTEM PROMPT:}} \\
Given a multi-hop question, its answer, and a set of supporting paragraphs,
determine the correct reasoning order that connects the paragraphs into a directed acyclic graph (DAG) matching the specified topology.
\\
\\
You must output ONLY a JSON object with a list of dictionaries where each dictionary has a "title" key and a "next" key.
The "next" key should contain the title of the next paragraph in the correct reasoning order. \\
The first title should be the starting point of reasoning, and the last title should be the \\
paragraph that leads to the final answer.
\\
\\
IMPORTANT: Output ONLY the JSON object, nothing else. Do not include any \\
explanation, markdown formatting, or extra text.
Example: if the reasoning order is P2 $\rightarrow$ P1, P3 $\rightarrow$ P1, P1 $\rightarrow$ P4, \
and the titles of P1, P2, P3, P4 are "Title1", "Title2", "Title3", "Title4" respectively, \\
then you should output: \\
\texttt{
    [{"title": "Title2", "next": "Title1"}, {"title": "Title3", "next": "Title1"}, 
    {"title": "Title1", "next": "Title4"}, {"title": "Title4", "next": "ANSWER"}]
}
\\
\midrule
\textbf{\textit{USER PROMPT:}} \\
\texttt{Graph type:} \textit{\{graph\_type\}} \\
\texttt{Number of supporting paragraphs:} \textit{\{num\_support\}} \\[4pt]
\texttt{Question:} \textit{\{question\}} \\
\texttt{Answer:} \textit{\{answer\}} \\[4pt]
\texttt{Supporting paragraphs (in random order):} \\
\quad\textit{\{paragraphs\_text\}} \\[4pt]
\texttt{Now, output the correct reasoning chain in the specified JSON format.} \\
    \bottomrule
    \end{tabular}
    \caption{Prompt template for validating evidence chain.}
    \label{tab:prompt_template_for_evidence_chain_validation}
\end{table}

\clearpage

\begin{table}[ht!]
\hsize\textwidth \linewidth\hsize
    \centering
    \begin{tabular}{p{15cm}}
    \toprule
    \textbf{Example Instantiated Prompt} \\
    \midrule
    \texttt{Graph type:} 2-hop \\
    \texttt{Number of supporting paragraphs:} 2 \\[4pt]
    \texttt{Question:} Who is the mayor of the city where the Battle of Gettysburg took place? \\
    \texttt{Answer:} John Smith \\[4pt]
    \texttt{Supporting paragraphs (in random order):} \\[2pt]
    \quad\texttt{--- Paragraph 1 ---} \\
    \quad\texttt{Title:} Gettysburg, Pennsylvania \\
    \quad\texttt{Text:} Gettysburg is a borough and the county seat of Adams County\ldots \\[2pt]
    \quad\texttt{--- Paragraph 2 ---} \\
    \quad\texttt{Title:} Battle of Gettysburg \\
    \quad\texttt{Text:} The Battle of Gettysburg was fought July 1--3, 1863, in and around the town\ldots \\[4pt]
    \texttt{Now, output the correct reasoning chain in the specified JSON format.} \\
    \midrule
    \textbf{Expected LLM Output:} \\
    \texttt{[\{``title'': ``Battle of Gettysburg'', ``next'': ``Gettysburg, Pennsylvania''\},} \\
    \quad\texttt{\{``title'': ``Gettysburg, Pennsylvania'', ``next'': ``ANSWER''\}]} \\
    \bottomrule
    \end{tabular}
    \caption{A concrete example of the fully instantiated prompt for a 2-hop question. The LLM receives the system-level instruction followed by the user prompt (\autoref{tab:prompt_template_for_evidence_chain_validation}), and is expected to produce the JSON output shown at the bottom.}
    \label{tab:evidence_chain_template_example}
\end{table}

\clearpage

\section{Iter-RetGen Prompt Templates}
\label{sec:iter_retgen_prompt_templates}
Below we illustrate prompt templates for generating embedding and Iter-RetGen. Templates for 2WikiMultihop and MuSiQue are identical, while for MultiHop-RAG we add source of content as many of its question decomposition revolve around this.
Following previous work~\citep{zhuang2024efficientragefficientretrievermultihop}, we adopt the same prompt template on four evaluate datasets for Iter-RetGen.

\begin{table}[h!]
    \centering
    \begin{tabular}{p{7.5cm}}
    \toprule
        \textbf{Document Chunk Prompt Template for 2WikiMultihop and MuSiQue} \\
    \midrule
        Title: \textit{[doc title]} \\
        Context: \textit{[doc text]} \\
    \bottomrule
    \end{tabular}
    \caption{Prompt template for generating embedding using BGE-m3 embedding model on 2Wiki and MuSiQue train and evaluate datasets.}
    \label{tab:embedding_prompt_template_2wiki_musique}
\end{table}

\begin{table}[ht!]
    \centering
    \begin{tabular}{p{7.5cm}}
    \toprule
        \textbf{Document Chunk Prompt Template for MultiHop-RAG} \\
    \midrule
        Title: \textit{[doc title]} \\
        Source: \textit{[doc source]} \\
        Context: \textit{[doc text]} \\

    \bottomrule
    \end{tabular}
    \caption{Prompt template for generating embedding using BGE-m3 embedding model on MultiHop-RAG evaluate dataset.}
    \label{tab:embedding_prompt_template_multihop_rag}
\end{table}

\clearpage

\begin{table}[ht!]
\hsize\textwidth \linewidth\hsize
    \setlength{\abovecaptionskip}{0.1cm}
    \setlength{\belowcaptionskip}{-0.5cm}
    \centering
    \footnotesize
    \begin{tabular}{p{15cm}}
    \toprule
        \textbf{Iter-RetGen Prompt Template for 2WikiMultihop, MuSiQue and MultiHop-RAG} \\
        \midrule
        You should think step by step and answer the question after <Question> based on given knowledge embraced with <doc> and </doc>. Your answer should be after <Answer> in JSON format with key "thought" and "answer", their value should be string. \\

        Here are some examples for you to refer to: \\
        <doc> \\
        \{\{KNOWLEDGE FOR THE QUESTION\}\} \\
        </doc> \\
        <Question>: In which year did the publisher of In Cold Blood form? \\
        Let's think step by step. \\
        <Answer>: \\
        \texttt{\`}\texttt{\`}\texttt{\`} json \\
        \{\{
            "thought": "In Cold Blood was first published in book form by Random House. Random House was form in 2001.",
            "answer": "2011"
        \}\} \\
        \texttt{\`}\texttt{\`}\texttt{\`} \\
        
        <doc> \\
        \{\{KNOWLEDGE FOR THE QUESTION\}\} \\
        </doc> \\
        <Question>: Who was in charge of the city where The Killing of a Sacred Deer was filmed? \\
        Let's think step by step. \\
        <Answer>: \\
        \texttt{\`}\texttt{\`}\texttt{\`} json \\
        \{\{
            "thought": "The Killing of a Sacred Deer was filmed in Cincinnati. The present Mayor of Cincinnati is John Cranley. Therefore, John Cranley is in charge of the city.",
            "answer": "John Cranley"
        \}\} \\
        \texttt{\`}\texttt{\`}\texttt{\`} \\
        
        <doc> \\
        \{\{KNOWLEDGE FOR THE QUESTION\}\} \\
        </doc> \\
        <Question>: Where on the Avalon Peninsula is the city that Signal Hill overlooks? \\
        Let's think step by step. \\
        <Answer>: \\
        \texttt{\`}\texttt{\`}\texttt{\`} json \\
        \{\{
            "thought": "Signal Hill is a hill which overlooks the city of St. John's. St. John's is located on the eastern tip of the Avalon Peninsula.",
            "answer": "eastern tip"
        \}\} \\
        \texttt{\`}\texttt{\`}\texttt{\`} \\
        
        Now based on the given doc, answer the question after <Question>. \\
        <doc> \\
        \{documents\} \\
        </doc> \\
        <Question>: \{question\} \\
        Let's think step by step. \\
        <Answer>: \\
    \bottomrule
    \end{tabular}
    \caption{Prompt template for Iter-RetGen on 2Wiki, MuSiQue and MultiHop-RAG evaluate datasets.}
    \label{tab:iter-retgen_prompt_template}
\end{table}

\clearpage

\section{Accuracy Prompt for GPT 5 Mini}
\label{sec:acc_prompt_gpt5mini}
Following previous work~\citep{zhuang2024efficientragefficientretrievermultihop}, we disclose accuracy prompt for evaluating end-to-end QA performance from \autoref{tab:e2e_treehop_vs_rag}.

\begin{table}[ht!]
    \centering
    \begin{tabular}{p{7.5cm}}
    \toprule

    \textbf{Accuracy Prompt} \\
    \midrule
    You are an experienced linguist who is responsible for
    evaluating the correctness of the generated responses.
    You are provided with question, the generated responses
    and the corresponding ground truth answer. Your task
    is to compare the generated responses with the ground
    truth responses and evaluate the correctness of the generated responses. Response in JSON format with key
    "response" and value "yes" or "no". \\
    Question: \textit{[question]} \\
    Prediction: \textit{[prediction]} \\
    Ground-truth Answer: \textit{[answer]} \\

    \bottomrule
    \end{tabular}
    \caption{Accuracy Prompt for Evaluating end-to-end QA performance}
    \label{tab:acc_prompt_gpt3.5}
\end{table}

\section{QA Prompt for End-to-End Evaluation}
\label{sec:qa_prompt_for_e2e_eval}
For the downstream LLMs (Llama3.1-8B-Instruct, Qwen3.6-27B and Qwen2.5-7B-Instruct) in our end-to-end QA performance evaluation, we applied default chat templates, directly attached questions to the user prompt, and adopted the same system prompt for the three LLMs below, where retrieved chunks are naively concatenated in descending order of cosine similarity, then further concatenated in retrieval iteration order.

For decoding, we set top-$p$, top-$k$ and repetition penalty to 1 and the maximum number of new tokens to 1024 for all three models, and set temperature to 0.8 for Llama3.1-8B-Instruct, 0.3 for Qwen3.6-27B and 0.6 for Qwen2.5-7B-Instruct.

\begin{table}[h!]
    \centering
    \begin{tabular}{p{7.5cm}}
    \toprule
    \textbf{QA Prompt} \\
    \midrule
    Answer the question based on the context below. Respond "Unsure about answer" if not sure about the answer. \\
Here is the context: \\
Title: \textit{[title1]} \\
\textit{[content1]} \\
\\
Title: \textit{[title2]} \\
\textit{[content2]} \\
\\
… \\

    \bottomrule
    \end{tabular}
    \caption{Downstream LLM Prompt for Evaluating end-to-end QA performance}
    \label{tab:qa_prompt_for_e2e_eval}
\end{table}

\clearpage

\section{Experiment on All Query Types}
\label{sec:experiment_all_query_types}
In \autoref{sec:eval_datasets}, we explicitly selected iterative retrieval questions for our benchmark, filtering out questions that do not require iterative retrieval to gather all necessary information.
However, some may raise concerns regarding TreeHop’s generalizability when questions that do not demand a retrieval system's multi-hop capability are also considered. To address this, we present a revision of benchmarks reported in \autoref{tab:retrieve_treehop_vs_rag}, where we test TreeHop's performance with all query types included. All experimental settings are identical to those described in \autoref{subsec:eval_metric}, except that comparison queries have been added back to the benchmarks. Note, that the results for the HotpotQA and MuSiQue datasets are omitted, as no question types were excluded from them in our original experiment.

\begin{table}[ht!]
\hsize\textwidth \linewidth\hsize
\centering
\begin{tabular}{lcccccccc}
\toprule

\multicolumn{1}{c}{} & \multicolumn{4}{c}{\textbf{2WIKI}} & \multicolumn{4}{c}{\textbf{MULTIHOP RAG}} \\

\multicolumn{1}{l}{\textbf{Retriever}} & \multicolumn{1}{c}{\textbf{Recall@5}} & \multicolumn{1}{c}{\textbf{Chain}} & \multicolumn{1}{c}{\textbf{Avg.$K$}} &\multicolumn{1}{c|}{\textbf{Latency}} & \multicolumn{1}{c}{\textbf{Recall@5}} & \multicolumn{1}{c}{\textbf{Chain}} & \multicolumn{1}{c}{\textbf{Avg.$K$}} &\multicolumn{1}{c}{\textbf{Latency}} \\

\midrule

\multicolumn{9}{l}{\textit{Baselines}} \\
$\text{Direct-R@5}$ & 57.5 & 29.4 & 5 & 0.003 & 54.6 & 36.3 & 5 & 0.002 \\

\midrule

\multicolumn{9}{l}{\textit{Advanced RAG}} \\

$\text{\textbf{Iter-RetGen}(8B)@5 iter2}$ & 66.4 & 57.8 & 9.9 & 4.713 & 60.3 & 40.6 & 9.9 & 4.888 \\
$\text{\textbf{Iter-RetGen}(8B)@5 iter3}$ & 69.8 & 60.9 & 14.8 & 7.286 & 63.5 & 42.9 & 14.8 & 7.357 \\
$\text{\textbf{Iter-RetGen}(0.6B)@5 iter2}$ & 63.1 & 53.6 & 6.4 & 2.996 & 58.7 & 39.4 & 9.5 & 2.904 \\
$\text{\textbf{Iter-RetGen}(0.6B)@5 iter3}$ & 65.0 & 56.5 & 8.8 & 4.790 & 59.4 & 43.0 & 13.7 & 4.695 \\
$\text{\textbf{EfficientRAG}@5}$ & 63.1 & 53.8 & 4.4 & 2.802 & 53.9 & 38.0 & 5.7 & 2.859 \\

\midrule

\multicolumn{9}{l}{\textit{Ours}} \\

$\text{\textbf{TreeHop}@5 iter2}$ & \textbf{69.2} & \textbf{61.0} & 9.2 & \textbf{0.103} & \textbf{61.6} & \textbf{43.0} & 7.7 & \textbf{0.115} \\
$\text{\textbf{TreeHop}@5 iter3}$ & \textbf{72.8} & \textbf{62.4} & 13.1 & \textbf{0.202} & \textbf{64.3} & \textbf{46.9} & 11.3 & \textbf{0.180} \\

\bottomrule
\end{tabular}

\caption{TreeHop performance when all query types are included. \textbf{Chain} is the complete evidence-chain recall defined in \autoref{subsec:eval_metric}. \textbf{Bold} numbers indicate the best performance in the same iteration among retrievers.}

\label{tab:treehop_performance_all_query_types}

\end{table}

Overall, the existence of comparison queries does not alter the conclusion in \autoref{sec:results} that TreeHop was both performant and efficient among advanced iterative retrieval approaches. Surprisingly, TreeHop’s ability to retrieve complementary information still improves recall for comparison queries, likely because the previous retrieval step sometimes fails to retrieve all necessary information.

\clearpage

\section{Comparison with Hierarchical Summarisation Retrieval}
\label{sec:raptor_comparison}
RAPTOR~\citep{sarthi2024raptor} indexes a corpus by recursively clustering and summarising passages, so retrieval draws on a pool holding both original passages and LLM-written summaries. We evaluate both of its modes under the protocol of \autoref{subsec:eval_metric}: \textit{collapse} flattens the tree into a single top-$K$, while \textit{traversal} descends layer by layer, taking top-5 at each. Only leaf nodes carry an evidence title, so only they can close a chain; at $K=5$ collapse spends 1.07--1.16 of its five slots on summary nodes, and traversal reaches leaves only at its final iteration.

\begin{table}[ht!]
\hsize\textwidth \linewidth\hsize
\footnotesize
\setlength{\tabcolsep}{2pt}
\begin{center}
 \setlength\tabcolsep{.75\tabcolsep}
\resizebox{\textwidth}{!}{%
\begin{tabular}{@{}lcccc|cccc|cccc|cccc@{}}
\toprule
 & \multicolumn{4}{c|}{\textbf{HOTPOT}} & \multicolumn{4}{c|}{\textbf{2WIKI}} & \multicolumn{4}{c|}{\textbf{MUSIQUE}} & \multicolumn{4}{c}{\textbf{MULTIHOP-RAG}} \\
\multicolumn{1}{l}{\textbf{Retriever}} & \multicolumn{1}{c}{\textbf{Recall}} & \multicolumn{1}{c}{\textbf{Chain}} & \multicolumn{1}{c}{\textbf{Avg.$K$}} & \multicolumn{1}{c|}{\textbf{Lat.}} & \multicolumn{1}{c}{\textbf{Recall}} & \multicolumn{1}{c}{\textbf{Chain}} & \multicolumn{1}{c}{\textbf{Avg.$K$}} & \multicolumn{1}{c|}{\textbf{Lat.}} & \multicolumn{1}{c}{\textbf{Recall}} & \multicolumn{1}{c}{\textbf{Chain}} & \multicolumn{1}{c}{\textbf{Avg.$K$}} & \multicolumn{1}{c|}{\textbf{Lat.}} & \multicolumn{1}{c}{\textbf{Recall}} & \multicolumn{1}{c}{\textbf{Chain}} & \multicolumn{1}{c}{\textbf{Avg.$K$}} & \multicolumn{1}{c}{\textbf{Lat.}} \\

\midrule

\multicolumn{17}{l}{\textit{Baselines}} \\
$\text{Direct-R@5}$ & 68.3 & 43.5 & 5 & 0.002 & 49.3 & 12.7 & 5 & 0.002 & 45.4 & 16.4 & 5 & 0.002 & 48.6 & 15.8 & 5 & 0.002 \\
$\text{Direct-R@10}$ & 71.2 & 50.5 & 10 & 0.003 & 53.2 & 16.9 & 10 & 0.003 & 49.8 & 20.2 & 10 & 0.002 & 53.0 & 19.3 & 10 & 0.002 \\

\midrule

\multicolumn{17}{l}{\textit{Hierarchical Summarisation}} \\
$\text{\textbf{RAPTOR}@5 collapse}$ & 65.2 & 38.8 & 5 & 5.438 & 47.1 & 10.9 & 5 & 5.640 & 42.5 & 14.1 & 5 & 5.231 & 47.5 & 14.5 & 5 & 6.372 \\
$\text{\textbf{RAPTOR}@10 collapse}$ & 72.4 & 50.2 & 10 & 5.489 & 51.5 & 15.7 & 10 & 5.910 & 50.8 & 21.9 & 10 & 5.165 & 54.8 & 20.4 & 10 & 6.309 \\
$\text{\textbf{RAPTOR}@15 collapse}$ & \textbf{76.0} & \textbf{56.0} & 15 & 5.347 & 53.6 & 18.3 & 15 & 5.721 & \textbf{55.2} & 26.2 & 15 & 5.004 & 56.2 & 22.6 & 15 & 6.256 \\
$\text{\textbf{RAPTOR}@5 traversal iter2}$ & 40.9 & 19.7 & 10 & 6.843 & 23.4 & 8.1 & 10 & 6.969 & 27.4 & 9.7 & 10 & 6.417 & 28.6 & 8.7 & 10 & 7.777 \\
$\text{\textbf{RAPTOR}@5 traversal iter3}$ & 21.6 & 8.4 & 15 & 7.095 & 11.0 & 4.2 & 15 & 7.259 & 18.0 & 5.6 & 15 & 6.647 & 10.1 & 5.2 & 15 & 8.419 \\

\midrule

\multicolumn{17}{l}{\textit{Ours}} \\
$\text{\textbf{TreeHop}@5 iter3}$ & 74.5 & 54.9 & 14.2 & \textbf{0.202} & \textbf{66.8} & \textbf{48.1} & 14.0 & \textbf{0.204} & 53.7 & \textbf{27.7} & 13.5 & \textbf{0.200} & \textbf{66.1} & \textbf{27.9} & 12.2 & \textbf{0.181} \\

\bottomrule
\end{tabular}%
}

\caption{RAPTOR against direct retrieval and TreeHop. \textbf{Bold} numbers indicate the best performance at a comparable Avg.$K$ ($\approx$14--15).}

\label{tab:retrieve_raptor_vs_treehop}

\end{center}
\end{table}

At a matched Avg.$K$ of 5 RAPTOR trails direct retrieval by 1.1\%-3.1\% recall and 1.3\%-4.7\% Chain, the summary slots costing it more than the enriched pool returns. A larger Avg.$K$ recovers this: at $K=15$, comparable to TreeHop's 12.2--14.2 chunks at the third iteration, RAPTOR reaches a Chain of 56.0\% on HotpotQA against TreeHop's 54.9\%, and 26.2\% on MuSiQue against 27.7\%. On 2WikiMultiHop and MultiHop-RAG it reaches only 18.3\% and 22.6\%, against 48.1\% and 28.9\%. The gap tracks how far the evidence chain must bridge: summarisation enriches the pool that the first passages are drawn from, but it never uses a retrieved passage to reformulate the query, so the later hops of a 2WikiMultiHop chain stay out of reach. Traversal is weaker on every dataset, since expansion is confined to children of already selected nodes and a passage becomes unreachable once its ancestor is pruned. Latency compounds the picture: collapse costs 5.0--6.4\,s per query and traversal, which searches once per layer, 6.4--8.4\,s, between 25 and 47 times TreeHop's at a comparable Avg.$K$. Restructuring the index therefore buys no efficiency advantage over rewriting the query, on top of requiring an LLM summarisation pass over every cluster at every layer to build.

\clearpage

\section{Experiment on Different Base Embedding Model and Prompt Template}
\label{sec:analysis_on_different_base_embedding_model}
TreeHop is fundamentally trained upon embeddings generated by BGE-m3. This raises question on whether the same model architecture can adapt to other base models and other prompt templates. To validate this, we replaced BGE-m3 with Qwen3-Embedding-4B model~\citep{yang2025qwen3technicalreport}, trained and benchmarked the TreeHop model variant using the identical train dataset, evaluate datasets and metrics demonstrated in~\autoref{sec:train_data_construction},~\autoref{sec:eval_datasets} and~\autoref{subsec:eval_metric}. All hyper-parameters are the same as in~\autoref{sec:model_training} except that the number of negatives is 6. Prompt templates for both BGE-m3 embedding and Qwen3-Embedding-4B are shown in ~\autoref{tab:prompt_template_on_different_base_embedding_model} results are shown in~\autoref{tab:retrieve_treehop_variant_vs_rag}.

As shown in \autoref{tab:retrieve_treehop_variant_vs_rag}, the TreeHop architecture transfers effectively to the Qwen3-Embedding-4B base model, yielding substantial improvements over both the direct retrieval baseline and the BGE-m3-based TreeHop variant across all four benchmarks.

First, comparing the two direct retrieval baselines, Direct-R(Qwen)@5 already outperforms Direct-R(BGE)@5 by a considerable margin (e.g., 80.8 vs.\ 68.3 on HotpotQA, 64.3 vs.\ 49.3 on 2Wiki), confirming that the stronger base embeddings of Qwen3-Embedding-4B provide a higher-quality starting point. Critically, TreeHop preserves and amplifies this advantage: TreeHop(Qwen)@5 at iteration 2 achieves 84.1 recall on HotpotQA and 81.0 on 2Wiki, representing absolute gains of 10.5 and 18.3 points over TreeHop(BGE)@5 iter2, respectively. On MuSiQue and MultiHop-RAG, the Qwen-based variant similarly leads by 7.4 and 6.1 points at the second iteration. These consistent improvements across datasets of varying difficulty indicate that TreeHop's iterative tree-expansion mechanism is not overfitted to BGE-m3's embedding space but generalizes to a different encoder architecture.

\begin{table}[ht!]
    \centering
    \begin{tabular}{p{7.5cm}}
    \toprule
    \textbf{Question Prompt for BGE-m3} \\
\midrule

    \textit{[question]} \\

\midrule

    \textbf{Document Chunk Prompt for BGE-m3} \\
\# Paragraph title: \textit{[title]} \\
\#\#\# Content \\
\textit{[content]} \\

\midrule

    \textbf{Question Prompt for Qwen3-Embedding-4B} \\
    Given a search query, retrieve relevant passages that answer the query. \\
    Here is the query: \textit{[question]} \\
\midrule
    \textbf{Document Chunk Prompt for Qwen3-Embedding-4B} \\
Title: \textit{[title]} \\
\textit{[content]} \\
… \\

    \bottomrule
    \end{tabular}
    \caption{Prompt templates for BGE-m3 and Qwen3-embedding-4B.}
    \label{tab:prompt_template_on_different_base_embedding_model}
\end{table}

Second, the iterative refinement behavior is preserved under the new base model. Moving iteration from the second to the third, TreeHop(Qwen)@5 continues to gain recall on all four benchmarks (e.g., +0.3\% on HotpotQA, +0.5\% on 2Wiki, +1.3\% on MuSiQue, +3.2\% on MultiHop-RAG), mirroring the pattern observed with the BGE variant.

Third, the latency profile reflects the expected cost of using a larger base encoder. TreeHop(Qwen)@5 iter2 incurs roughly 0.16-0.17s per query compared to 0.11s for TreeHop(BGE)@5 iter2, an approximately 1.5 times increase attributable to the 4B-parameter Qwen embedder (embeds sentences with 2560 dimensions) versus the lighter BGE-m3 (embeds sentences with 1024 dimensions). This trade-off is consistent across datasets and iterations, and the absolute latency remains practical for offline or batch retrieval scenarios. Users can therefore choose between the BGE variant for latency-sensitive deployments and the Qwen variant when recall is the primary objective.

\newpage

\begin{table}[b!]
\hsize\textwidth \linewidth\hsize

\small
\setlength{\tabcolsep}{2pt}
\begin{center}
 \setlength\tabcolsep{.75\tabcolsep}
\begin{tabular}{@{}lccc|ccc|ccc|ccc@{}}
\toprule

 & \multicolumn{3}{c|}{\textbf{HOTPOT}} & \multicolumn{3}{c|}{\textbf{2WIKI}} & \multicolumn{3}{c|}{\textbf{MUSIQUE}} & \multicolumn{3}{c}{\textbf{MULTIHOP-RAG}} \\
\multicolumn{1}{l}{\textbf{Retriever}} & \multicolumn{1}{c}{\textbf{Recall}} & \multicolumn{1}{c}{\textbf{Avg.$K$}} & \multicolumn{1}{c|}{\textbf{Latency}} & \multicolumn{1}{c}{\textbf{Recall}} & \multicolumn{1}{c}{\textbf{Avg.$K$}} & \multicolumn{1}{c|}{\textbf{Latency}} & \multicolumn{1}{c}{\textbf{Recall}} & \multicolumn{1}{c}{\textbf{Avg.$K$}} & \multicolumn{1}{c|}{\textbf{Latency}} & \multicolumn{1}{c}{\textbf{Recall}} & \multicolumn{1}{c}{\textbf{Avg.$K$}} & \multicolumn{1}{c}{\textbf{Latency}}\\
 
\midrule
\multicolumn{12}{l}{\textit{Baselines}} \\
$\text{Direct-R(BGE)@5}$ & 68.3 & 5 & 0.002 & 49.3 & 5 & 0.002 & 45.4 & 5 & 0.002 & 48.6 & 5 & 0.002 \\
$\text{Direct-R(Qwen)@5}$ & 80.8 & 5 & 0.002 & 64.3 & 5 & 0.002 & 54.5 & 5 & 0.002 & 57.9 & 5 & 0.002 \\

\midrule

\multicolumn{9}{l}{\textit{Ours}} \\ 

$\text{\textbf{TreeHop}(BGE)@5 iter2}$ & 73.6 & 9.5 & 0.117 & 62.7 & 9.5 & 0.114 & 52.0 & 9.2 & 0.113 & 60.2 & 8.7 & 0.108 \\ 
$\text{\textbf{TreeHop}(BGE)@5 iter3}$ & 74.5 & 14.2 & 0.202 & 66.8 & 14.0 & 0.204 & 53.7 & 13.5 & 0.200 & 66.1 & 12.2 & 0.181 \\ 

$\text{\textbf{TreeHop}(Qwen)@5 iter2}$ & 84.1 & 9.3 & 0.167 & 81.0 & 9.4 & 0.160 & 59.4 & 9.1 & 0.160 & 66.3 & 8.2 & 0.163 \\ 
$\text{\textbf{TreeHop}(Qwen)@5 iter3}$ & 84.4 & 13.9 & 0.282 & 81.5 & 13.9 & 0.281 & 60.7 & 13.4 & 0.263 & 69.5 & 11.3 & 0.246 \\ 

\bottomrule
\end{tabular}

\caption{Results of baselines, concurrent advanced RAG solutions and TreeHop variant on four MHQA datasets. Direct-R@5 denotes the direct retrieval at top-5 baseline retrieval setting (See \autoref{subsec:eval_metric} for baseline and advanced RAG details).}

\label{tab:retrieve_treehop_variant_vs_rag}

\end{center} 
\end{table}

\clearpage

\section{Analysis on The Size of Train Dataset}
\label{sec:analysis_on_the_size_of_training_dataset}
To explore the possible room of further improving TreeHop by increasing the size of training dataset, we have randomly sampled our current training dataset and trained TreeHop upon 75\%, 50\% and 25\% of the training dataset, respectively. Throughout the experiment, we adopted the environment and hyper-parameters introduced in ~\autoref{sec:model_training}.

\begin{table}[ht!]
\hsize\textwidth \linewidth\hsize
\centering
\begin{tabular}{lcccccc}
\toprule

& \multicolumn{2}{c}{\textbf{2WIKI}} & \multicolumn{2}{c}{\textbf{MUSIQUE}} & \multicolumn{2}{c}{\textbf{MULTIHOP RAG}} \\

\multicolumn{1}{l}{\textbf{Subsamble Size}} & \multicolumn{1}{c}{\textbf{Recall@K}} & \multicolumn{1}{c}{\textbf{K}} & \multicolumn{1}{c}{\textbf{Recall@K}} & \multicolumn{1}{c}{\textbf{K}} & \multicolumn{1}{c}{\textbf{Recall@K}} & \multicolumn{1}{c}{\textbf{K}} \\

\midrule

\multicolumn{1}{l}{Direct-R@5} & 49.3 & 5 & 45.4 & 5 & 48.6 & 5 \\

\midrule

\multicolumn{7}{l}{\textit{100\% of Training Dataset}} \\

\multicolumn{1}{l}{\textbf{TreeHop}@5 iter2} & 62.7 & 9.5 & 52.0 & 9.2 & 60.2 & 8.7 \\

\multicolumn{1}{l}{\textbf{TreeHop}@5 iter3} & 66.8 & 14.0 & 53.7 & 13.5 & 66.1 & 12.2 \\

\midrule
\multicolumn{7}{l}{\textit{75\% of Training Dataset}} \\

\multicolumn{1}{l}{\textbf{TreeHop}@5 iter2} & 61.8(0.9\textit{$\downarrow$}) & 9.4 & 48.8(3.2\textit{$\downarrow$}) & 9.1 & 59.3(0.9\textit{$\downarrow$}) & 8.5 \\

\multicolumn{1}{l}{\textbf{TreeHop}@5 iter3} & 65.6(1.2\textit{$\downarrow$}) & 13.7 & 49.7(4.0\textit{$\downarrow$}) & 13.3 & 63.6(2.5\textit{$\downarrow$}) & 11.9 \\

\midrule
\multicolumn{7}{l}{\textit{50\% of Training Dataset}} \\

\multicolumn{1}{l}{\textbf{TreeHop}@5 iter2} & 60.4(1.4\textit{$\downarrow$}) & 9.2 & 47.0(1.8\textit{$\downarrow$}) & 8.9 & 56.0(3.3\textit{$\downarrow$}) & 7.8 \\

\multicolumn{1}{l}{\textbf{TreeHop}@5 iter3} & 62.3(3.3\textit{$\downarrow$}) & 11.9 & 47.2(2.5\textit{$\downarrow$}) & 12.4 & 58.8(4.8\textit{$\downarrow$}) & 10.9 \\

\midrule
\multicolumn{7}{l}{\textit{25\% of Training Dataset}} \\

\multicolumn{1}{l}{\textbf{TreeHop}@5 iter2} & 57.1(3.3\textit{$\downarrow$}) & 8.4 & 46.0(1.0\textit{$\downarrow$}) & 7.9 & 53.1(2.9\textit{$\downarrow$}) & 6.6 \\

\multicolumn{1}{l}{\textbf{TreeHop}@5 iter3} & 59.3(3.0\textit{$\downarrow$}) & 11.4 & 46.4(0.8\textit{$\downarrow$}) & 11.3 & 55.6(3.2\textit{$\downarrow$}) & 8.7 \\

\bottomrule
\end{tabular}

\caption{TreeHop performance trained upon subsampled dataset.}

\label{tab:subsample_train_dataset_size}

\end{table}

Our experiment reveals that TreeHop's performance decays as the training dataset size decreases. We observe a notable improvement on all benchmarks, indicating that there is still room for improvement if we further scale up the training dataset.

\clearpage

\section{Why do we need Query Rewriters Like TreeHop for Multi-hop Question Answering?}
\label{why_need_query_rewriter_for_MHQA}
Though there exists various query rewriters~\citep{zhuang2024efficientragefficientretrievermultihop,shao2023enhancingretrievalaugmentedlargelanguage,vakulenko2020question,ma2023queryrewritingretrievalaugmentedlarge,vakulenko2020question}, readers may still raise doubt in the necessity of adopting query rewriters for multi-hop question answering (MHQA).
At first glance, one could argue that simply retrieving more documents should be sufficient:
a larger candidate set is more likely to contain the facts needed for a multi‑step answer,
and the iterative refinement that TreeHop performs would appear unnecessary.
we argue that query rewriters is still necessary to tasks like MHQA, and there are two pivotal aspects support our belief.

\paragraph{Role of Prompt Length}
After retrieval, the selected passages are concatenated with the original question and fed to a LLM for answer generation. A larger number of retrieved chunks induces extremely lengthy prompt that could easily elicit LLM’s lost-in-the-middle syndrome. Therefore, the number of retrieved documents $K$ is also critical to the accuracy and efficiency. Even we have tools like reranker models, they do not preclude irrelevant context. Rewriters like TreeHop connect semantically distant documents without extremely large number of $K$.


\paragraph{Queries that are Intrinsically Unanswerable by a Single Retrieval}
Consider the question: \textit{“What province shares a border with the province where Lago District is located?”} A naïve, one‑shot retrieval with the original query will only return documents that mention Lago District (e.g., its location, demographics, etc.). None of these passages contain the name of the bordering province, because that information resides in a different document that is only reachable after we first learn that Lago District belongs to Niassa Province. A query rewriter reads the initial passage, infer the intermediate entity, and issue a second, more specific query such as:
\textit{“Which provinces border Niassa Province?”}. The second retrieval step is now targeted at the missing piece of the reasoning chain, dramatically increasing the probability of finding the final answer.

This two‑step reasoning pattern is typical of real‑world user queries, where all the required facts cannot be gathered without multi-step reasoning. Without an intermediate rewriting stage, the system would either have to retrieve an impractically large $K$, (thereby inflating the prompt) or fail to retrieve the necessary evidence altogether.

On a higher level, the overarching goal of retrieval system is to reduce the information entropy in the prompt by minimizing irrelevant retrieved documents while still reaching a high recall, while direct retrieval often introduces huge number of irrelevant information with the recall. Query rewriters address this problem by re‑formulating the query to target the next missing piece of knowledge, thereby steering the retrieval toward more focused, low-entropy results.

\clearpage

\section{Intuition Behind TreeHop's Architecture}
\label{intuition_behind_treehop_architecture}
To clarify the underlying intuition in \autoref{sec:model_architecture}, let us revisit the example presented in the introduction: For the question \textit{“Who is the grandfather of Donald Trump?”}, suppose the first retrieval yields a chunk stating \textit{“Donald Trump’s father is Fred Trump.”} Traditional systems use LLMs to rewrite the query as \textit{“Who is the father of Fred Trump?”} for the next hop. Our TreeHop operates at the embedding level to achieve this rewrite dynamically, thereby achieves efficiency. Under the hood, the model is fed with query-chunk embedding pair, it then replaces \textit{“grandfather of Donald Trump”} information in the query embedding with \textit{“father of Fred Trump”} information in the retrieved chunk embedding. The $ForgetGate$ term in the \autoref{eq:treehop_architecture1} serves for this replacing purpose by attending over the query-chunk pair and subtracting the query semantics that the chunk has already resolved. Meanwhile, $UpdateGate$ is adopted to selectively integrate new information from the retrieved chunk to form the next query embedding (\autoref{eq:treehop_architecture1}). The $UpdateGate$ is a cross attention module that selectively maintain only information that is necessary for the next retrieval from chunk embedding (\autoref{eq:treehop_architecture2}), e.g., \textit{"Fred Trump"}.

\section{Multilingual Capability of TreeHop}
\label{multilingual_capability_of_treehop}
Since TreeHop is trained on embeddings produced by BGE-m3 and, during the inference phase, is also fed with query and chunk embeddings outputted by BGE-m3 (rather than raw text), we believe that TreeHop's multilingual capability is inherited from BGE-m3. This is because if two queries written in different languages are semantically close, BGE-m3 will output two similar embeddings, which will be close to each other in terms of cosine distance. Consequently, if the two query embeddings are similar, TreeHop will produce identical next query embeddings given the same chunk embedding.
TreeHop works exclusively in the embedding level,
it doesn't see the language difference; it only sees the semantic patterns produced by BGE-m3,
which are language-agnostic.

\section{More Explanation About Layer-Wise Top-K Pruning}
\label{more_explanation_to_layerwise_topk_pruning}
To begin with, $K$ is a hyper-parameter that controls the number of chunks to be retrieved given a query. However, this could result in an explosion in number of retrieved chunks in multi-hop retrieval: During initial retrieval step, a user query is vectorized and $K$ chunks with the highest similarities to the query are retrieved. Subsequently, TreeHop would generate $K$ embeddings for each of the query-chunk pair. In the next retrieval, we would retrieve top-$K$ chunks again for \textbf{every} generated embedding, resulting in a total of $K^2$ retrieved chunks. Generally, given the number of retrieval iterations $r$, the chunks grow exponentially as $K^r$, most of which are irrelevant. To address this overhead, at each retrieval iteration, our Layer-Wise Top-$K$ Pruning retains only the top-$K$ chunks by cosine similarity to the generated query embeddings, pruning the rest. This reduces the branching factor, prevents redundant retrievals and minimizes prompt overload for downstream LLMs.


\clearpage

\section{Case Study on TreeHop's Retrieval}
\label{sec:case_study_treehop_retrieval}
To enhance readers' understanding on how TreeHop works, below we present a retrieval case from the 2WikiMultiHop benchmark, with the retrieval settings being: two‑hop query, top‑5 retrieval. The cosine distances are directly obtained from the KNN search procedure presented by the Fiass library~\citep{johnson2019billion}.
For readability we use a hierarchical chunk identifier: “Chunk i.j” means chunk j retrieved in hop i (rank j among the top‑5).

\renewenvironment{framed}[1][\hsize]
  {\MakeFramed{\hsize#1\advance\hsize-\width \FrameRestore}}%
  {\endMakeFramed}

\begin{table}[b!]
\hsize\textwidth \linewidth\hsize
\footnotesize
\begin{framed}[\textwidth]

\textbf{Query:} \textit{"What is the place of birth of the director of the film Beowulf \& Grendel?"}

\vspace{+3mm}

\paragraph{The first retrieval iteration (by direct retrieval framework, BGE-m3)}

\begin{center}
    \begin{tabular}[0.98\textwidth]{lllc}
    \toprule
        \textbf{Rank} & \textbf{Annotation} & \textbf{Retrieved Chunk} & \textbf{Cosine Distance} \\

    \midrule

        \multirow{2}{*}{\textbf{1}} & \multirow{2}{*}{\textit{True Positive}} & \multicolumn{1}{p{9cm}}{"Beowulf \& Grendel is a 2005 Canadian‑Icelandic fantasy‑adventure film directed by \textbf{Sturla Gunnarsson} …"} & \multirow{2}{*}{0.071} \\

    \midrule

        \textbf{2} & - & "Wrath of Gods …, directed by Jon Gustafsson …" & 0.073 \\

    \midrule

        \multirow{2}{*}{\textbf{3}} & \multirow{2}{*}{-} & \multicolumn{1}{p{8.6cm}}{"Keith Bearden (born …) is an American screenwriter and director …"} & \multirow{2}{*}{0.077} \\

    \midrule

        \multirow{2}{*}{\textbf{4}} & \multirow{2}{*}{-} & \multicolumn{1}{p{8.6cm}}{"Wolf Gremm (26 Feb 1942 – 14 Jul 2015) was a German film director …"} & \multirow{2}{*}{0.080} \\

    \midrule

        \multirow{2}{*}{\textbf{5}} & \multirow{2}{*}{-} & \multicolumn{1}{p{8.6cm}}{"Werner Herzog (born 5 Sep 1942) is a German film director …"} & \multirow{2}{*}{0.081} \\

    \bottomrule
    \end{tabular}

\end{center}

\vspace{+2mm}
\paragraph{The second retrieval iteration (by TreeHop framework)}

\begin{center}
\begin{tabular}{lllc}
    \toprule
        \textbf{Rank} & \textbf{Annotation} & \textbf{Retrieved Chunk} & \textbf{Cosine Distance} \\

    \midrule

        \multirow{3}{*}{\textbf{1.1}} & \multirow{3}{*}{\textit{True Positive}} & \multicolumn{1}{p{9cm}}{"Sturla Gunnarsson (born 30 Aug 1951) is an Icelandic‑Canadian film and television director and producer… \textbf{Gunnarsson was born in Reykjavík in 1951}.""} & \multirow{3}{*}{1.531} \\

    \midrule

        \multirow{2}{*}{\textbf{1.2}} & \multicolumn{1}{p{2cm}}{Layer‑wise top‑5 pruned} & \multicolumn{1}{p{9cm}}{"Ágúst Guðmundsson (born 29 Jun 1947) is an Icelandic film director and screenwriter."} & \multirow{2}{*}{1.689} \\

    \midrule

        \multirow{2}{*}{\textbf{1.3}} & \multicolumn{1}{p{2cm}}{Layer‑wise top‑5 pruned} & \multicolumn{1}{p{9cm}}{"Ágúst Guðmundsson studied French and Icelandic in Reykjavík..."} & \multirow{2}{*}{1.709} \\

    \midrule

        \multirow{2}{*}{\textbf{1.4}} & \multicolumn{1}{p{2cm}}{Redundancy pruned} & \multicolumn{1}{p{9cm}}{"Beowulf \& Grendel is a 2005 Canadian‑Icelandic fantasy‑adventure film directed by Sturla Gunnarsson..."} & \multirow{2}{*}{1.716} \\

    \midrule

        \multirow{2}{*}{\textbf{1.5}} & \multicolumn{1}{p{2cm}}{Layer‑wise top‑5 pruned} & \multicolumn{1}{p{9cm}}{"Keith Bearden (born in Middletown, Connecticut) is an American screenwriter and director."} & \multirow{2}{*}{1.729} \\

    \midrule

        \multirow{4}{*}{\textbf{2.1}} & \multirow{4}{2cm}{Redundancy pruned} & \multicolumn{1}{p{9cm}}{"Wrath of Gods (2006 documentary) tells the story of the dramatic circumstances Canadian director Sturla Gunnarsson and his crew had to go through during the making of the film ‘Beowulf \& Grendel’."} & \multirow{4}{*}{1.583} \\

    \midrule

        \multirow{2}{*}{\textbf{2.2}} & \multicolumn{1}{p{2cm}}{Redundancy pruned} & \multicolumn{1}{p{9cm}}{"Beowulf \& Grendel is a 2005 Canadian‑Icelandic fantasy‑adventure film directed by Sturla Gunnarsson."} & \multirow{2}{*}{1.627} \\

\end{tabular}

\end{center}
\end{framed}
\end{table}
We highlighted the true-positive chunks that actually contain pertinent information in the annotation column. Other annotations indicate why a chunk is not expanded further by our proposed stopping criterion:
\begin{itemize}
    \item \textbf{Redundancy‑pruned}: the chunk has already appeared in the current or a previous hop.
    \item \textbf{Layer‑wise Top‑K pruned:} the chunk’s cosine similarity falls outside the top‑5 for that hop.
\end{itemize}

\clearpage

\begin{table}[ht!]
\hsize\textwidth \linewidth\hsize
\footnotesize
\begin{framed}[\textwidth]
\begin{center}
\begin{tabular}[0.98\textwidth]{lllc}

    \midrule

        \multirow{3}{*}{\textbf{2.3}} & \multirow{3}{2cm}{Redundancy pruned} & \multicolumn{1}{p{9cm}}{"Sturla Gunnarsson (born 30 Aug 1951) is an Icelandic‑Canadian film and television director and producer… Gunnarsson was born in Reykjavík in 1951."} & \multirow{3}{*}{1.674} \\

    \midrule

        \multirow{2}{*}{\textbf{2.4}} & \multicolumn{1}{p{2cm}}{Layer‑wise top‑5 pruned} & \multicolumn{1}{p{9cm}}{"Mikael Håfström (born 1 Jul 1960) is a Swedish film director and screenwriter."} & \multirow{2}{*}{1.729} \\

    \midrule

        \multirow{2}{*}{\textbf{2.5}} & \multicolumn{1}{p{2cm}}{Layer‑wise top‑5 pruned} & \multicolumn{1}{p{9cm}}{"Ágúst Guðmundsson studied French and Icelandic in Reykjavík and filmmaking at the National Film School in London."} & \multirow{2}{*}{1.733} \\

    \midrule

        \multirow{2}{*}{\textbf{3.1}} & \multicolumn{1}{p{2cm}}{Redundancy pruned} & \multicolumn{1}{p{9cm}}{"Keith Bearden (born in Middletown, Connecticut) is an American screenwriter and director."} & \multirow{2}{*}{1.538} \\

    \midrule

        \multirow{2}{*}{\textbf{3.2}} & \multirow{2}{2cm}{Layer‑wise top‑5 pruned} & \multicolumn{1}{p{9cm}}{"Mike Binder (born 2 Jun 1958) is an award‑winning American film director, screenwriter, producer and actor."} & \multirow{2}{*}{1.670} \\

    \midrule

        \multirow{2}{*}{\textbf{3.3}} & \multicolumn{1}{p{2cm}}{Layer‑wise top‑5 pruned} & \multicolumn{1}{p{9cm}}{"Edward Bernds (12 Jul 1905 – 20 May 2000) was an American screenwriter and director..."} & \multirow{2}{*}{1.706} \\

    \midrule

        \multirow{2}{*}{\textbf{3.4}} & \multicolumn{1}{p{2cm}}{Layer‑wise top‑5 pruned} & \multicolumn{1}{p{9cm}}{"Seth Gordon (born 15 Jul 1976) is an American film director, producer, screenwriter and film editor."} & \multirow{2}{*}{1.773} \\

    \midrule

        \multirow{2}{*}{\textbf{3.5}} & \multicolumn{1}{p{2cm}}{Layer‑wise top‑5 pruned} & \multicolumn{1}{p{9cm}}{"David Frankel (born 2 Apr 1959) is an American film director, screenwriter and producer."} & \multirow{2}{*}{1.781} \\

    \midrule

        \multirow{2}{*}{\textbf{4.1}} & \multicolumn{1}{p{2cm}}{Redundancy pruned} & \multicolumn{1}{p{9cm}}{"Wolf Gremm (26 Feb 1942 – 14 Jul 2015) was a German film director and screenwriter."} & \multirow{2}{*}{1.507} \\

    \midrule

        \multirow{2}{*}{\textbf{4.2}} & \multicolumn{1}{p{2cm}}{Layer‑wise top‑5 pruned} & \multicolumn{1}{p{9cm}}{"Franz Josef Gottlieb (1 Nov 1930 – 23 Jul 2006) was an Austrian film director and screenwriter."} & \multirow{2}{*}{1.719} \\

    \midrule

        \multirow{2}{*}{\textbf{4.3}} & \multirow{2}{2cm}{Layer‑wise top‑5 pruned} & \multicolumn{1}{p{9cm}}{"Rainer Werner Fassbinder (31 May 1945 – 10 Jun 1982), sometimes credited as R.W. Fassbinder, was a West German filmmaker..."} & \multirow{2}{*}{1.730} \\

    \midrule

        \multirow{3}{*}{\textbf{4.4}} & \multirow{3}{2cm}{Layer‑wise top‑5 pruned} & \multicolumn{1}{p{9cm}}{"Hans Billian (born Hans Joachim Hubert Backe, 15 Apr 1918 in Breslau (now Wrocław, Poland) – 18 Dec 2007 in Gräfelfing, Bavaria) was a German film directo..."} & \multirow{3}{*}{1.730} \\

    \midrule

        \multirow{2}{*}{\textbf{4.5}} & \multirow{2}{2cm}{Redundancy pruned} & \multicolumn{1}{p{9cm}}{"Werner Herzog (born 5 Sep 1942) is a German film director, screenwriter, author, actor and opera director."} & \multirow{2}{*}{1.733} \\

    \midrule

        \multirow{2}{*}{\textbf{5.1}} & \multirow{2}{2cm}{Redundancy pruned} & \multicolumn{1}{p{9cm}}{"Werner Herzog (born 5 Sep 1942) is a German film director, screenwriter, author, actor and opera director."} & \multirow{2}{*}{1.341} \\

    \midrule

        \multirow{3}{*}{\textbf{5.2}} & \multirow{3}{2cm}{-} & \multicolumn{1}{p{9cm}}{"Rainer Werner Fassbinder (31 May 1945 – 10 Jun 1982), sometimes credited as R.W. Fassbinder, was a West German filmmaker..."} & \multirow{3}{*}{1.629} \\

    \midrule

        \multirow{2}{*}{\textbf{5.3}} & \multirow{2}{2cm}{Redundancy pruned} & \multicolumn{1}{p{9cm}}{"Wolf Gremm (26 Feb 1942 – 14 Jul 2015) was a German film director and screenwriter."} & \multirow{2}{*}{1.734} \\

    \midrule

        \multirow{2}{*}{\textbf{5.4}} & \multirow{2}{2cm}{Layer‑wise top‑5 pruned} & \multicolumn{1}{p{9cm}}{"Georg Tressler (25 Jan 1917 – 6 Jan 2007) was a Vienna‑born German film actor and film director."} & \multirow{2}{*}{1.737} \\

    \midrule

        \multirow{2}{*}{\textbf{5.5}} & \multirow{2}{2cm}{Layer‑wise top‑5 pruned} & \multicolumn{1}{p{9cm}}{"Werner W. Wallroth (28 Feb 1930 – 9 Aug 2011) was a German film director and screenwriter."} & \multirow{2}{*}{1.744} \\

    \bottomrule
\end{tabular}
\end{center}
\end{framed}
\end{table}

In conclusion, the true‑positive chunks (1 and 1.1) are retrieved at ranks 1 and 1.1 during the first and second retrieval iterations, respectively, confirming that TreeHop effectively propagates the query embedding and refines the retrieval in the second hop. At the second iteration, the TreeHop framework retrieves chunks 1.1 and 5.2, while all other chunks are filtered out by the proposed stopping criterion. Notably, although redundant chunks 2.1, 2.2, and 5.1 exhibit lower cosine distances, they are disregarded because layer-wise ranking is performed after removing redundant chunks.

\clearpage

\section{TreeHop Inference Procedure}
\label{sec:treehop_inference_procedure}
The full multi-hop inference loop, including both pruning rules of
\autoref{sec:stop_criterion}, is given in \autoref{fig:treehop_inference_step}.

\begin{figure}[th!]

\begin{framed}
\textbf{Input:} Initial user query text $x$, embedding model $f(\cdot)$ that produces embeddings for input text, total number of hops $N\ge1$, retriever $g(\cdot)$ that takes one query embedding and outputs top $K$ text chunks $\{c^i \}_{i=1}^K$ and respective embeddings $\{\upsilon^i\}_{i=1}^K$, cosine similarity scores $\{s^i\}_{i=1}^K$. \\ \\
\textbf{Output:} Retrieved text chunk set $\mathcal{C}$.

\vspace{+0.6em}
\begin{algorithmic}[1]
\STATE
$\mathcal{C} = \emptyset$ \\
$\mathcal{Q} = \{f(x)\}$    // Query embedding set \\

\FOR {$r  \in \{1, \dots, N \}$} 
    \STATE $\mathcal{S} = \emptyset$ \\
    \FOR {$q$ in $\mathcal{Q}$}
        \STATE {// Retrieve and iterate over top $K$ chunks, embeddings and similarity scores}
        \FOR {$c$, $\upsilon$, $s$ in $g(q_i, K)$}
            \IF {$c \notin \mathcal{C}$ }
                \STATE{// Include only distinct chunks.} \label{eq:redundancy_pruning} \\
                $\mathcal{S} \gets [q, c, v, s]$
            \ENDIF
        \ENDFOR

    \ENDFOR

    \STATE {$\mathcal{Q} = \emptyset$}

    $t = \textbf{TopSimilarityScore}(\mathcal{S}, K)$ // Get layer-wise $K$-th similarity score $t$ from set $\mathcal{S}$.
    \FOR {$q, c,\upsilon,s$ in $\mathcal{S}$}
        \IF {$s \ge t$}
            \STATE {$\mathcal{C} \gets c$} \label{eq:layerwise_top_pruning}\\
            $\mathcal{Q} \gets \textbf{TreeHop}(q, v)$
        \ENDIF
    \ENDFOR
\ENDFOR

\end{algorithmic}
\end{framed}
\vspace{-4mm}
\caption{Multi-hop inference steps and rule-based stopping criterion for TreeHop.}
\label{fig:treehop_inference_step}

\end{figure}

\end{document}